\documentclass[authoryear,3p,twocolumn,10pt]{elsarticle}

\usepackage{graphicx}
\usepackage{epsfig}

\usepackage{natbib}
\usepackage{amsmath, amsthm,amssymb}
\usepackage{lscape} % Wes Custom
\usepackage{epstopdf}
\usepackage{longtable}

\newcommand{\All}{\emph{TNO} }
\newcommand{\Cold}{\emph{Cold} }
\newcommand{\Hot}{\emph{Hot} }
\newcommand{\Close}{\emph{Close} }
\newcommand{\ColdGood}{\emph{Cold}\ensuremath{_{F}} }
\newcommand{\HotGood}{\emph{Hot}\ensuremath{_{F}} }

\newcommand{\aj}{AJ}           
\newcommand{\nat}{Nature}
\newcommand{\apj}{ApJ}

\newcommand{\apjl}{ApJL}
\newcommand{\pasp}{PASP}

\newcommand{\planss}{PSS}
\newcommand{\pasj}{pasj}

\begin{document}

\begin{frontmatter}

\title{The Luminosity Function of the Hot and Cold Kuiper belt Populations}

\tnotetext[t1]{Accepted by Icarus August 2nd, 2010.}

\author[1]{Wesley C. Fraser \corref{cor1}}
\ead{fraserw@gps.caltech.edu}

\author[1]{Michael E. Brown}

\author[1]{Megan E. Schwamb}

\address[1]{Division of Geological and Planetary Sciences, MS150-21, California Institute of Technology, 1200 E. California Blvd. Pasadena, CA 91101 USA }

\begin{abstract}
We have performed an ecliptic survey of the Kuiper belt, with an areal coverage of 8.9 square degrees to a 50\% limiting magnitude of $r'_{Sloan}=24.7$, and have detected 88 Kuiper belt objects,  roughly half of which received follow-up one to two months after detection. Using this survey data alone, we have measured the luminosity function of the Kuiper belt, thus avoiding any biases that might come from the inclusion of other observations. We have found that the \Cold population defined as having inclinations less than $5^o$ has a luminosity function slope $\alpha_{\Cold}=0.82\pm0.23$, and is different from the \Hot population, which has inclinations greater than $5^o$ and a luminosity function slope $\alpha_{\Hot}=0.35\pm0.21$. As well, we have found that those objects closer than 38 AU have virtually the same luminosity function slope as the \Hot population. This result, along with similar findings of past surveys demonstrates that the dynamically cold Kuiper belt objects likely have a steep size distribution, and are unique from all of the excited populations which have much shallower distributions. This suggests that the dynamically excited population underwent a different accretion history and achieved a more evolved state of accretion than the cold population. As well, we discuss the similarities of the \Cold and \Hot populations with the size distributions of other planetesimal populations. We find that while the Jupiter family comets and the scattered disk exhibit similar size distributions, a power-law extrapolation to small sizes for the scattered disk cannot account for the observed influx of comets. As well, we have found that the Jupiter Trojan and \Hot populations cannot have originated from the same parent popuation, a result that is difficult to reconcile with scattering models similar to the NICE model. We conclude that the similarity between the size distributions of the \Cold population and the Jupiter Trojan population is a striking coincidence.
\end{abstract}

\begin{keyword}
Kuiper Belt
Centaurs
Trojan asteroids
Accretion
\end{keyword}

\end{frontmatter}

\section{Introduction}
The size distribution  is one of the most fundamental properties of a planetesimal population. As the size of an object is primarily determined from its accretion and collisional disruption histories, the size distribution can reveal important information on the accretion and collisional history of that population \citep[for a recent example, see][]{Bottke2005b}.

Unlike the closer populations such as the asteroid belt, whose proximity has allowed the accurate measurement of their size distributions \citep[see for example][]{Jedicke1998,Jewitt2000}, the distance to the Kuiper belt has prevented an equally detailed determination of its size distribution. Observations thus far have demonstrated that for the Kuiper belt as a whole, the size distribution for objects with diameters, $D\gtrsim200$ km,  is well described by a power-law. At some size $50\lesssim D \lesssim 150$ km, the size distribution rolls-over to a shallower distribution \citep{Bernstein2004, Fuentes2009, Fraser2009}. Smaller than the roll-over, the size distribution appears to remain shallow to objects as small as $D\sim1$ km \citep{Schlichting2009}.

While the accuracy of current measurements prevents a detailed modelling of the history of objects in the region, some insight has already been gained.  The steepness of the large object size distribution implies that for this population, accretion was a short-lived phenomenon, likely not more than a few 100 Myr \citep{Gladman2001,Kenyon2002,Fraser2008}. The paucity of the observed belt, and the existence of the largest known members demonstrates that the belt has undergone significant mass depletion, losing as much or more than 99\% of its primordial mass \citep{Stern1997a,Kenyon1998,Jewitt2002,Fuentes2009}. The existence of the roll-over at sizes larger than $D\gtrsim 50$ km suggests that the belt has undergone significant collisional comminution in a region of significantly increased density compared to today \citep{Kenyon2004,Benavidez2009,Fraser2009c}.

As well, recent observations have suggested that the size distribution of those dynamically cold (low inclinations and eccentricities) Kuiper belt objects is different than that of the dynamically hot (large inclinations and eccentricities) population \citep{Levison2001}. \citet{Bernstein2004} found that the size distribution of objects with inclinations, $i < 5^o$ was steeper than the size distribution of objects with $i>5^o$. \citet{Fuentes2008} found similar results showing that the cold population had a steeper slope than the mixed populations as a whole. This result implies that the hot and cold populations are genetically separate populations, that have significantly different accretion and evolution histories.

The results of \citet{Bernstein2004} were drawn from samples of objects compiled from many different surveys. This practice was necessitated by available data. All surveys had either shallow limiting magnitudes and a large areal coverage, or vice-versa. The result was that most surveys did not have a sufficient range of objects for which the size distribution could be accurately measured from that survey alone. The practice of using data from multiple surveys opens the results to the possibility of being affected by  calibration issues and variations in sky density which could lead to an incorrect measurement of the size distribution. When these effects were properly accounted for, \citet{Fraser2008,Fraser2009} found that these data could not be used to reliably test differences in subpopulations of the Kuiper belt.

Here we present the results of a new Kuiper belt survey. By virtue of the survey's design, a large number of objects were discovered and followed to determine their inclinations, over a range of sizes sufficient to measure the size distribution for the hot and cold population without the use of other surveys.  In Section~\ref{sec:observations} we present our observations, and our data reductions and discovery techniques. In Section~\ref{sec:analysis} we present the analysis of our results, in Section~\ref{sec:discussion} we discuss the implications of our results, and we end with concluding remarks in Section~\ref{sec:conclusions}.

\section{Observations and Reductions} \label{sec:observations}

\subsection{Discovery Observations}
The discovery observations were made on October 18, 2009 (UT) with Suprime-cam on the 8.2 m Subaru telescope \citep{Miyazaki2002}. Suprime-cam is 10-chip mosaic camera, with a field-of-view of roughly 34'$\times$27' with $\sim15^{"}$ chip-gaps, and has a pixel scale of 0.2". The observations consisted imaging 34 fields, each visited three times in the r' filter with the camera long axis oriented horizontally in RA, with each visit consisting of a single 200 s exposure. The first and last images of a field were separated by roughly 6.5 hours. The observed fields were all within $\sim1.5^o$ of the ecliptic and within $\sim10^o$ of opposition at the discovery epoch. The total areal coverage of our survey after accounting for chip-gaps was 8.93 square degrees. Details of the individual fields are shown in Table~\ref{tab:pointings} and Figure~\ref{fig:pointings}. The observations included multiple images of the D1 and D4 fields of the Canada-France-Hawaii Telescope Supernova Legacy Survey \citep{Astier2006} at various airmasses to provide both photometric and astrometric calibrations of the discovery images.

The images were pre-processed with standard techniques; the bias levels were removed, and a master bias frame, set as the average of 10 bias images was removed from all science frames. A master sky-flat was produced from the science frames using a clipped median filter, and was removed from the science images producing background variations no larger than $\sim1\%$ across a mosaic.

Photometric calibrations were done on a chip-by-chip basis using the Mega-pipe source catalog of the D1 and D4 fields \citep{Gwyn2008}. The zeropoint was found to be identical for all chips within uncertainties, with a value, $Z_{r_{Subaru}}=27.50\pm0.02$. The conversion from the Suprime-cam instrumental magnitudes to those of Mega-pipe was found to be 

\begin{equation}
m(r')_{Suprime} = m(r')_{Mega} - 0.018(g'_{Mega}-r'_{Mega}).
\label{eq:Color_megaSubaru}
\end{equation}
\noindent
Atmospheric extinction was measured from the science frames. The airmass correction was found to be 0.08 magnitudes per unit increase in airmass. Comparison of common background sources in each triplet revealed no large seeing or transparency variations throughout the night (see Figures~\ref{fig:seeing} and \ref{fig:airmass}); the night was photometric.

The Suprime-cam field of view is highly distorted, especially near field edges, requiring calibration to ensure accurate astrometry of moving sources in the observations. Scamp \citep{Bertin2006} was used to measure second order spatial distortions of the Suprime-cam field, from the D1 and D4 fields on a chip-by-chip basis. Little variance in the distortions was found over the airmass range of these observations. Thus, a master distortion map for each chip was produced from all images of the D1 and D4 fields. Absolute calibration of each science image was done using the USNO-B catalog, and the master distortion map, resulting in residuals of roughly 0.3". Relative astrometry between the three images of each field - a triplet - was performed by matching sources between an image, and the reference image, chosen as the image of the triplet with the lowest absolute residuals. This resulted in an excellent common astrometric solution for the triplet, with residuals of $\lesssim 0.06"$ between all three images. This calibrations procedure ensured reliable astrometry to both maximize the discovery efficiency of moving sources, as well as secure the possibility of follow-up on future dates.

\subsection{Moving Object Search}
The Suprime-cam data were utilized to search for moving objects in the Solar system, in and beyond the orbits of the gas-giant planets. To characterize the detection efficiency of our moving object detection method, artificial moving point sources were implanted in the observations. 

The stellar point-spread function (PSF) was generated on an image-by-image and chip-by-chip basis from 10-15 hand-selected, visually inspected point sources for each chip. It was found that a spatially constant PSF was sufficient to model the stellar shape across a chip. The PSFs were created using the tools in the \emph{doaphot} package of \emph{IRAF} and for each image consisted of an average Moffat profile with look-up table of the PSF stars of that chip.

Random artificial sources were generated on Sun-bound orbits, with semi-major axes, $18\leq a\leq1000$ AU, eccentricities, $e\leq 0.6$, and inclinations, $i\leq 90^o$. Anomaly, periapse, and nodal angles were chosen at random to ensure that the sources fell in the images and could be either approaching or receding from their nodes. 20-30 artificial sources were planted in each chip with apparent magnitudes in the range $21\leq r' \leq 26.5$ with image-to-image flux variations matching those measured from 30-40 bright stars common to each image of a triplet, thus ensuring that any extinction variations throughout the observations were accounted for. If an artificial source's motion would cause it to drift more than 0.2 pixels during an exposure, it was broken up into multiple fainter sources with 0.2 pixel spacing and a total flux equal to that of the artificial source. Thus, trailing effects were fully accounted for in our search.

Moving objects were identified by their motions between images of a triplet. Using Sextractor \citep{Bertin2002}, all sources in an image were tabulated. Stationary sources common to each image in a triplet were flagged, and ignored from further consideration, leaving only those sources who were either  moving,  or had highly variable fluxes. The remaining sources of an image triplet were then searched for motion consistent with Sun-bound objects. Candidate sources were chosen  as any three individual point-source detections, one per image, whose centroid moved at least one pixel (0.2") and at most $\sim$5" between each image. These candidate sources were then further filtered by \emph{fit\_radec} \citep{Bernstein2000}. If the best-fit orbit to the candidate had a chi-squared larger than 7.5, the candidate was rejected. The results of this were approximately 16000 candidate sources which were then visual inspected by an operator for final acceptance or rejection.

The majority of visually inspected candidates were false, primarily caused by cosmic ray impacts, or extended sources which were identified by Sextractor as more than one source in some images. Of the 16,000 candidates, roughly a third were matched with implanted artificial sources. From these sources, the performance of the search routine was characterized. The results are shown in Figure~\ref{fig:efficiency}. As can be seen, the detection efficiency is well described by the familiar functional form

\begin{equation}
\eta(\mbox{r'})=\frac{A}{2}\left(1-\tanh{\left(\frac{r'-m_*}{g}\right)}\right)
\label{eq:efficiency}
\end{equation}

\noindent
where $A$, $m_*$, and $g$ are the maximum efficiency, half-maximum magnitude, and half-width parameters. The survey achieved a 50\% limiting magnitude for discovery of r'=24.7. The search was found to be sensitive to objects as distant as $\sim 900$ AU, beyond which, insufficient motion was exhibited for reliable detection. No significant variation in search efficiency with inclination or eccentricity was found (see Figures ~\ref{fig:eff_i} and \ref{fig:eff_r}).

The search resulted in the detection of 88 sources. The sources' distances and inclinations were determined with \emph{fit\_radec}. Photometry was measured using standard aperture photometry techniques with corrections for airmass applied (see Figure~\ref{fig:airmass}). Aperture corrections were measured from the bright stars from which the PSFs were generated. Photometric uncertainties were determined from the function

\begin{equation}
\Delta r' = \delta + \gamma 10^{\frac{r'-Z}{2.5}}
\label{eq:delta-r}
\end{equation}

\noindent
where $\delta$ is a constant representing the extinction stability of the night in question, $\gamma$ is a constant proportional to the parameters of the detector \citep[see][for a derivation]{Fraser2008}, and $Z$ is the telescope photometric zeropoint. This function was fit to the difference in measured and planted magnitudes of the artificial sources. The best-fit parameters are $(\delta,\gamma)=(0.028,3.25)$. The best-fit $\delta$ is in agreement with the scatter in extinction exhibited in Figure~\ref{fig:airmass}. Along with the measured fluxes with uncertainties sampled per measurement from Equation~\ref{eq:delta-r}, these results are presented in Table~\ref{tab:objects}. 

One detected object was identified as previously known Plutino 2004 VT75 which was not known to be in the observed fields prior to detection. This detection confirmed the accuracy of both the photometric and astrometric calibrations. The detected position of 2004 VT75 was $\sim 0.2"$ from that predicted from the Minor Planet Center. As well, this source was detected across different chips demonstrating that the astrometric fits for different chips of the same mosaic image are reliable in an absolute sense.

\subsection{Follow-up}
For a subset of the objects, follow-up observations were performed with the Low Resolution Imaging Spectrometer (LRIS) on the Keck-I telescope on the nights of November 21st, 22nd and December 21st, 2009 (UT). LRIS is a dual channel (blue and red) camera, with each camera, a 2 CCD mosaic having approximately a 6X8' field of view, with a $\sim25^{"}$ gap, and a pixel scale of $0.135^"$. This particular instrument is convenient for our follow-up observations, as the positional uncertainty ellipse of our targets at the time of follow-up was slightly smaller than the areal coverage of a single detector CCD. 

The follow-up observations consisted of pairs of images at the positions of each target, predicted from \emph{fit\_radec}, with temporal spacing between each image sufficient to detect the object's motion. By virtue of the dual channel imaging abilities of the camera,  images  in both the g and r LRIS filters were taken simultaneously. Exposure times were scaled from the flux of each source in the discovery images, and were chosen such that each source would have a photometric signal-to-noise ratio of $\sim10$ in each image. Images of the D1 and D4 Supernova fields were taken for calibration purposes.

Due to time constraints, only those objects who's distance might place them beyond 39 AU, and with magnitude in discovery images brightward of $r'=24.5$ were followed. 41 objects satisfied these constraints and were pursued in the follow-up observations. 

Due to the geometry of the follow-up observations, asteroid confusion was a concern. The predicted rate of motion of the follow-up targets was used as a diagnostic of whether or not a detected moving object was the correct source. The second CCD of the LRIS detector provided a measure of this asteroid confusion rate, as no Kuiper belt objects were expected to fall within this region of the images. In the November follow-up data, 3 asteroids with brightnesses and rates of motion consistent with our Kuiper belt targets were found in the second CCD for all follow-up exposures. With the exception of two follow-up pointings, our objects were not confused, as only one moving target per field was identified with a rate of motion consistent with the targeted source. This is consistent with our asteroid confusion rate determined from the other CCD. Special attention for the two objects with confused follow-up was made during the December observations. From the december data, the interloper in the November observations was easily rejected achieving reliable links between the discovery and follow-up images for all follow-up targets. No confusion occurred during the December follow-up.

Our follow-up efforts were rewarded with 100\% success rate for all 41 targets satisfying our distance and magnitude cuts. The conditions on the November nights were photometric and of moderate seeing, with full-width at half maximum $\lesssim 1.5^"$. On the night in December, while it was photometric, the seeing conditions were mediocre, reducing the observing efficiency compared to the November nights. As such, all of the 41 objects for which we attempted follow-up have month-long arcs, while only a small subset of 10 have arcs longer than this. The results of the follow-up efforts are presented in Table~\ref{tab:objects}.

Astrometric and photometric calibrations of the LRIS images were performed using the same techniques as applied to the Suprime-cam images. The astrometric distortions of the LRIS field were small, but noticeable, and varied with airmass. Distortion maps extracted from the calibration images at airmasses similar to the science images provided residual errors of $\sim0.3 -0.4"$ with respect to the USNO-B catalog.

For the LRIS blue-channel, the photometric zeropoint was found to be $Z_{g_{Keck}}=27.91\pm0.03$ with an airmass term $A_{g}=0.15\pm0.05$. During the observations, the LRIS red-channel detectors were plagued by electronics difficulties causing charge transfer inefficiencies that were dependent on source flux. This effect was found to be as much as $\sim10\%$ in engineering images. As such, we adopt a 10\% uncertainty on the zeropoint of this detector, and find, $Z_{r_{Keck}}=28.55\pm0.1$. The red-airmass term was found to be $A_{r}=0.08\pm0.03$.

The conversions between the LRIS r and g magnitudes and the r' and g' Mega-pipe magnitudes were found to be

\begin{equation}
m(r)_{Keck}=m(r'_{Mega})-0.2 (g'_{Mega}-r'_{Mega})
\label{eq:r_megaKeck}
\end{equation}

\noindent
and

\begin{equation}
m(g)_{Keck}=m(g'_{Mega})+1.2(g'_{Mega}-r'_{Mega}).
\label{eq:g_megaKeck}
\end{equation}

\noindent
The relatively large correction between $g_{Keck}$ and $g'_{Mega}$ is caused by the LRIS beam-splitter configuration which, in the configuration we utilized, truncated the red side of the $g_{Keck}$ filter while preserving the throughput across the $r_{Keck}$ filter.

\subsection{Color Conversion}
To facilitate the combination of the discovery and follow-up images, all photometry needed to be converted to a single system. The Sloan system \citep{Smith2002} was chosen. The conversion from Megaprime magnitudes to the Sloan system are

\begin{equation}
r'_{Mega} = r'_{Sloan}-0.024 (g'_{Sloan}-r'_{Sloan})
\label{eq:r_megaSloan}
\end{equation}

\noindent
and

\begin{equation}
g'_{Mega} = g'_{Sloan}-0.153 (g'_{Sloan}-r'_{Sloan}).
\label{eq:g_megaSloan}
\end{equation}

\noindent
From Equations~\ref{eq:Color_megaSubaru}, \ref{eq:r_megaKeck}, and \ref{eq:g_megaKeck} and the conversions presented above, we find that the conversions from the Subaru and Keck photometry to the Sloan system are

\begin{equation}
r'_{Sloan} = r_{Keck}-0.094 (g_{Keck}-r_{Keck}),
\label{eq:r_keckSloan_r}
\end{equation}

\begin{equation}
g'_{Sloan} = g_{Keck}-0.445 (g_{Keck}-r_{Keck}),
\label{eq:g_keckSloan_g}
\end{equation}

\noindent
and

\begin{equation}
r'_{Sloan} = r_{Subaru}-0.019 (g_{Keck}-r_{Keck}).
\label{eq:r_subaruSloan_r}
\end{equation}

Using Equations~\ref{eq:r_keckSloan_r}-\ref{eq:r_subaruSloan_r}, all photometry have been converted to the Sloan system. The photometry presented in Table~\ref{tab:objects} are the weighted averages of all measurements presented here. When no $(g_{Keck}-r_{Keck})$ colour was available, the average color of all those objects that did receive follow-up, $(g_{Keck}-r_{Keck})=1.21$ was used to convert to the Sloan survey. As the range of Keck colors for our sample is $0.68\leq (g_{Keck}-r_{Keck}) \leq 1.96$, we estimate the error in the conversion caused by using this average color is $\sim0.02$ magnitudes, negligibly small.

In Figure~\ref{fig:colour-inc} we present the $(g'_{Sloan}-r'_{Sloan})$ color versus inclination for those objects with follow-up observations. From this figure, it can be seen that our sample exhibits similar behaviour to the Kuiper belt objects in the Minor Planet Center analyzed by \citet{Peixinho2008}. That is, there is a correlation with colour and inclination; low inclination objects are typically redder than higher inclination objects. As both our colours and inclinations are significantly more inaccurate than those in the sample analyzed by \citet{Peixinho2008}, a repeat of the same analysis on our data is not warranted, as no more insight would be gained than by visually comparing our sample to theirs (see Figure 1 of \citet{Peixinho2008}).

\section{Analysis} \label{sec:analysis}
As found by \citet{Bernstein2004}, their seems to be some variation in the Kuiper belt size distribution with orbital excitation. More specifically, they defined two populations, the \emph{Cold} sample, which contains those objects with heliocentric distances, $38<d<55$ AU and inclinations, $i<5^o$, and the \emph{Excited} population as all those Kuiper belt objects in the same distance range, but with large inclinations. \citet{Bernstein2004} found that for bright objects $(R<24)$, the \emph{Cold} population had a much steeper size distribution than the \emph{Excited} population. This result was found from a sample of Kuiper belt objects compiled from a large number of different surveys, providing a result possibly affected by calibration errors and sky density variations.

\citet{Elliot2005} analyzed the results of the Deep Ecliptic Survey, and found similar results. That is, that the cold classical objects have a steeper size distribution than the resonant, or excited populations. This result however, was drawn from a survey without a calibrated detection efficiency, making the result untrustworthy.

\citet{Fuentes2008} performed a similar analysis as \citet{Bernstein2004}, but with additional survey data. Their findings were similar, ie., that the hot population size distribution is shallower than that of the cold population. But much like \citet{Bernstein2004}, they considered data from multiple surveys and may have been affected by the same calibration errors and sky density variations.

\citet{Fraser2008,Fraser2009} reanalyzed the survey data considered in \citet{Bernstein2004} along with new observations, and introduced a technique which accounted for possible errors induced from combining the results of different surveys when measuring the size distribution. They found that the apparent differences between the hot and cold populations could not be disentangled from possible effects caused by the combination of different surveys. When correctly accounting for these effects, no statistical difference between the hot and cold population was found. Population differences as large as those inferred by \citet{Bernstein2004} might exist, but their presence could not be discerned from the available data sets.

The survey we present here, has sufficient depth, and number of detections to probe the differences in size distributions of the hot and cold populations alone, without the need for other data. By doing so, we can ensure that the biases that may be introduced from the use of multiple datasets are avoided here providing a clean and reliable test for any differences between the hot and cold populations of the Kuiper belt.

\subsection{The Samples \label{sec:subsamples}}
We want to test the findings of \citet{Bernstein2004}. To that end, we define subsets of our detections, which are similar to those considered by \citet{Bernstein2004}. They are defined as follows:

\begin{enumerate}
\item We define \All sample, as all the detections in our sample.
\item Like \citet{Bernstein2004} we define the \Cold population as those objects with $38<d<55$ AU and $i<5^o$. We also further refine this, and define the \ColdGood sample as only those sources from the \Cold sample who received follow-up observations in addition to the discovery observations.
\item We define the \HotGood and \Hot samples as those objects with  $38<d<55$ AU and $i>5^o$, who have and have not received follow-up observations respectively.
\item We define the \Close population, as those objects with $30<d<38$. We do not make any distinction about follow-up, as the majority of the \Close population was not observed beyond the discovery observations.
\end{enumerate}

We note here that the exact inclination division between the \Cold and \Hot populations is uncertain. Measurements of the inclination distributions of the \Hot and \Cold samples suggest that the vast majority of cold classical objects will have inclinations less than $i\sim5^o$ \citep{Brown2001,Gulbis2010}. Analysis of the optical colour distribution of Kuiper belt objects reveals a correlation of colour with inclination \citep[for a recent example, see]{Peixinho2008} that suggests there is a division between low-inclination red objects, and low-inclination neutral objects at $i\sim12^o$. Given the uncertainty of the inclination division that separates the \Cold and \Hot populations, we consider the full range of inclination divisions testable by our observations and primarily discuss the historical division of $i=5^o$ unless otherwise stated.

The \Close population is a mix of objects in mean-motion resonances with Neptune, the Centaurs, and the scattered disk objects. On the other hand, the \Hot population consists of hot classical, resonant, and scattered disk objects. While the exact fractional mix of each of these populations cannot be determined from our observations, the Minor Planet Center can guide us as to which populations dominate these subsamples.  Analysis of the objects in the Minor Planet Center, near opposition in late October, within a few degrees of the ecliptic, reveals 13 objects satisfying $30<d<38$ AU, 6 of which are Plutinos, and 36 objects satisfying $d>38$ AU and $i>5^o$, of which only a few are Plutinos, with the majority being hot classical and scattered disk objects. As the relative mixes of each dynamical class are different for the \Hot and \Close subsamples, any differences detected in the size distributions of the subsamples can provide insight into the size distributions of the underlying populations.

\subsection{The Methods \label{sec:methods}}
As the Kuiper belt objects we detect are not resolved, we have no measure of their sizes, and therefore cannot measure the size distribution of our sample directly. Rather, we must use an in-direct technique, and infer the size distribution from the luminosity function. Brightward of $R\sim25$, the luminosity function is well represented by a power-law. That is, the differential surface density of objects with magnitude $m$ per square-degree is given by 

\begin{equation}
\Sigma(m)=\alpha \ln(10) 10^{\alpha (m-m_o)}
\label{eq:LF}
\end{equation}

\noindent
where $\alpha$ is the logarithmic slope of the power-law, and $m_o$ is the magnitude at which there is one object per square degree with that magnitude, or brighter. Recent measurements suggest that, for the \All sample, $\alpha=0.73-0.76$ and $m_o\sim 23.4$ in the R-band \citep{Fuentes2009,Fraser2009}. It is worth noting that the luminosity function has a break, or transition from  the steep slope for large objects, to a shallower distribution for fainter sources. The transition occurs around $R\sim 25$, which is at the tail end of our survey sensitivity. As such, the break will have little to no effect on our observations, and we ignore it in our analysis.

It can be shown that if the size distribution of a population is a power-law, with slope $q$, ie., $\frac{dN}{dr}\propto r^{-q}$, then the luminosity function of that population is given by Equation~\ref{eq:LF}, with slope $\alpha=\frac{q-1}{5}$. As such, the slopes of the size distributions of each population can be determined from their luminosity functions \citep[see][for a thorough discussion of this technique]{Fraser2008}.

The fits of Equation~\ref{eq:LF} is done with a Bayesian maximum likelihood technique. Specifically, the likelihood

\begin{eqnarray}
L\left(\left\{m\right\} | \alpha, m_o\right) &=& e^{-\Omega \int \eta(m) \Sigma(m|\alpha,m_o) dm} \nonumber \\
	&& \times \prod_i \sigma(m_i) \Sigma(m_i|\alpha, m_o) \nonumber \\
\end{eqnarray}

 \noindent
 is maximized over the luminosity function slope and normalization parameters, $\alpha$  and $m_o$. Here $\Omega$ is the areal coverage of our survey, $\{m\}$ is the set of magnitudes for our discovered objects, with $m_i$ the magnitude of object $i$. $\sigma(m_i)$ is a functional representation of the uncertainty in the magnitude $m_i$. As standard practice, we adopt a gaussian representation, with widths equal to the uncertainty in the observed magnitude of each source.
 
To avoid potential errors induced at low detection efficiencies, we consider only those sources with probability of detection greater than 50\% and truncate our detection efficiency below this point. \citet{Fraser2008} has demonstrated that truncating the efficiency at 50\% as we do here causes the best-fit slope $\alpha$ to appear roughly 3-5\% steeper than in reality. This effect however, is much smaller than the uncertainty in the best-fit slopes we present. For the \ColdGood and \HotGood samples, our follow-up is limited to those sources with $r'_{Subaru}<24.5$. Thus $r'=24.5$ is our magnitude cut for the \ColdGood and \HotGood samples. In Equation ~\ref{eq:LF}  we utilize only the magnitudes from discovery observations alone, as utilizing the data from other filters could possibly introduce biases into the results. We note however, that when utilizing all available flux measurements, the change in the results was significantly smaller than the uncertainties in the best-fit parameters.

To test the quality of the best-fits, we utilize the Anderson-Darling statistic,

\begin{equation}
\Delta=\int_0^1 \frac{\left(S(m)-P(m)\right)^2}{P(m)\left(1-P(m)\right)}dP(m)
\label{eq:Anderson-Darling}
\end{equation}

\noindent
where $P(m)$ is the cumulative probability of detecting an object with magnitude $\leq m$, and $S(m)$ is the cumulative distribution of detections. We calculate the probability, $P(\Delta>\Delta_{obs})$ of finding a value, $\Delta$, larger than that of the observations, $\Delta_{obs}$, given the best fit-parameters by bootstrapping the statistic, ie., randomly drawing a subsample of objects from the best-fit power-law, with number equal to that detected in a particular sample, fitting Equation~\ref{eq:LF} to that random sample, and computing $\Delta$. Values of $P(\Delta>\Delta_{obs})$ near 0 indicate that the functional form is a poor representation of the data. 

\subsection{The Results}
The luminosity functions of the \All, \Cold, \Hot, and \Close samples are presented in Figure~\ref{fig:LF}. The results of the fits of Equation~\ref{eq:LF} to the various subsamples are shown in Figure~\ref{fig:contours}, and Table~\ref{tab:best-fits}.

As can be seen from Figures~\ref{fig:LF} and \ref{fig:contours}, the \ColdGood exhibits a much steeper luminosity function, with slope $\alpha_{Cold}=0.82\pm0.23$ than the \HotGood population which has slope $\alpha_{Hot}=0.35\pm0.21$. The slopes of these two populations differ at more than the 1-$\sigma$ level. Both of these samples are well described by power-laws, as exhibited by their Anderson-Darling statistics (see Table~\ref{tab:best-fits}). Similar slopes within the uncertainties are found from the \Hot and \Cold samples. But these values are less trustworthy as the inaccurate inclinations of those objects which did not receive follow-up cause an uncertain amount of mixing between the two subsamples.

In addition, the \Close sample exhibits virtually the same luminosity function slope, $\alpha_{Close}=0.40\pm0.15$ as the \HotGood ( and \Hot) sample. As the \Close and \Hot subsamples are made up of different fractions of the Centaur, resonant, hot classical, and scattered populations, these results suggests that all of these populations exhibit equally shallow luminosity functions. If one or more of the populations had a luminosity function as steep as the \Cold population, it would likely be detectable as a difference in $\alpha$ between the \Hot and \Close populations, the steeper being the subsample with the greatest fraction of the steep dynamical population. This however, is not apparent. Indeed, the slope does not change when considering the \Hot and \Close populations together, resulting in $\alpha_{Hot+Close}=0.40\pm0.12$.

The luminosity function of \All sample has a slope $\alpha_{TNO}=0.56\pm0.1$, which is significantly shallower than the slope $\alpha\sim 0.75$ found from other recent measurements in the same magnitude range \citep{Fuentes2009, Fraser2009}. This surprising result can be understood relatively simply. The \All sample, is a mixture of the \Hot, and \Cold populations, which our results have shown have significantly different slopes. As such, we expect the \All sample to have a slope bounded by the slopes of the \Hot $(\alpha_{Hot}\sim0.4)$ and \Cold $(\alpha_{Cold}\sim0.8)$ populations. Our observations were made at a ecliptic longitude where plutinos, which are primarily members of the hot population, are preferentially at perihelion (see above). As such, we expect a greater fraction of Plutinos detected in our observations, than at other longitudes where Plutinos are not at perihelion. As the \Hot population has a shallower luminosity function than the \Cold population, we expect to see a flatter luminosity function for the \All sample at this longitude than at others. This simple reasoning can account for the variation in luminosity function slopes seen in past surveys \citep[see][]{Fraser2008}.

Along with the size distribution differences between the \Hot and \Cold populations,  the colour distribution \citep[][ and see Figure~\ref{fig:colour-inc}]{Tegler2000,Trujillo2002,Peixinho2008}, inclination distribution \citep{Brown2001}, binary fraction \citep{Stephens2006} and lack of large objects on dynamically cold orbits \citep{Brown2008}  suggest that these two populations are different. There is no clear evidence however, for a clean separation in inclination between the two. Rather, it is more likely that a smooth gradient exists, and as such, the historical division of $i_{div}=5^o$ chosen to separate these two populations is relatively arbitrary. As such, we measured the luminosity function slopes while varying $i_{div}$. The results are presented in Figure~\ref{fig:alpha-inc}.

As can be seen, the \Cold and \Hot population slopes become significantly different for $i_{div}\gtrsim 5^o$. When considering all objects, not just those with follow-up, the slope of the \Hot population seems to become a constant, with $\alpha_{Hot}\sim0.4$ for $i_{div}\geq6^o$. This suggests that above $6^o$ the fraction of the population with a steep size distribution is insignificant, and the observed objects come almost entirely from the population with a shallow size distribution. As the uncertainty on  $\alpha$ is still quite large, these observations however, are insufficient to confirm such a hypothesis.

To test the significance of the apparent difference in luminosity functions of the \Hot and \Cold populations, we utilize two different statistical tests. We first utilize the Kuiper-variant of the Kolmogorov-Smirnov test as a non-parametric measure of the significance in the observed difference between two populations \citep{Press2002}. We randomly bootstrapped from the cumulative luminosity function of the \Cold sample, a sample of objects equal in number to the \Hot sample. We then calculated the KS-statistic of this random sample compared to the \Cold sample. This process was repeated, and from this the probability of finding a simulated KS-statistic as large or larger than that found from the actual \Hot population was determined. We found that the two populations were drawn from separate parent populations at the 80-90\% significance levels for inclination divisions $i_{div} \geq 4^o$.

While the KS-test provides a non-parametric test, a better measure of the significance difference of the \Hot and \Cold populations can be found with the knowledge that the two populations are well described by power-laws. To that end, we also utilized the Anderson-Darling statistic, as done in \citet{Bernstein2004,Fraser2008}. Following similar procedures as above, by randomly sampling from the best-fit power-law of the \ColdGood sample we determined the probability of finding a simulated Anderson-Darling statistic as large or larger than that found from the actual \HotGood population as compared to the observed \ColdGood sample. We found that the two populations were drawn from separate parent populations at greater than the 3-$\sigma$ significance for inclination divisions $i_{div} \geq 4^o$.

\section{Discussion} \label{sec:discussion}

\subsection{Implications for the Kuiper belt}
The observed luminosity functions imply size distribution slopes $q_{\Cold}=5.1\pm1.1$, $q_{\Hot}=2.8\pm1.0$, and $q_{\Close}=3.0\pm0.8$ for the \Cold, \Hot, and \Close populations. The consistency of the \Hot and \Close populations leaves us to consider the two populations together, implying a slope for the combined population of $q_{\Hot+\Close}=3.0\pm0.6$.

Though the results presented by \citet{Bernstein2004} and \citet{Fuentes2008} were possibly affected by sky density variations and calibration errors, we find similar conclusions, ie., the \Cold population size distribution \emph{is} steeper than the \Hot population size distribution. Given the similarity in these repeated measurements of the Kuiper belt luminosity function, the results stand confident.

While over the magnitude range of our observations, both the \Hot and \Cold populations are well described by power-law models, other works have put into question whether the power-law behaviour extends to all brightnesses. The fact that the luminosity function of the \All population exhibits a roll-over at magnitudes $R\gtrsim25$ is well accepted. The results of \citet{Bernstein2004} however, suggest that the power-law behaviour of the \Cold population breaks down for the brightest objects. Specifically they find that at the bright-end, the \Cold population luminosity function has a steeper slope than at fainter magnitudes. Similarly, \citet{Morbidelli2009} suggests that the \Hot and \Cold populations are intimate mixtures of two primordial populations, one with a steep size distribution, and one with a shallow size distribution. In such a scenario, both populations should exhibit shallow slopes for the largest objects, then turn-up to a steeper slope for smaller objects, before finally breaking to a shallow slope at some size where collisional processing has dominated. To that end, we test the power-law behaviour of both the \Hot and \Cold populations.

For the \Cold population, a simple extrapolation of the best-fit luminosity function suggests that the brightest object on the sky in this population should have a magnitude of $R\sim19.5$. The entire low-latitude Kuiper belt has been surveyed for objects to a limiting magnitude of $R\lesssim21$ \citep{Trujillo2003}, and the brightest known \Cold object has $R\sim21.3$, nearly a full magnitude fainter. This observation confirms the results of \citet{Bernstein2004}. Namely, that the power-law behaviour of the \Cold population luminosity function cannot extend to the brightest targets. Rather the power-law must be truncated. The exact behaviour cannot be determined from the observations we present here. The suggestion by \citet{Morbidelli2009} that the cold population size distribution will have a shallow slope for the largest objects, and a turn-up to the observed steep is excluded by this result.

Similarly, extrapolating the best-fit power-law luminosity function of the \Hot population suggests that this population is well described by a power-law to its brightest objects. Indeed, when excluding the largest objects which are known to have different albedos than the objects observed here \citep{Stansberry2008} the absolute magnitude distribution of the biggest objects has a slope compatible with our observations \citep{Brown2008,Morbidelli2008}. To test the exactness of a power-law requires combination of other survey data sensitive to the brightest \Hot members. As well this analysis should include modelling of the latitude distribution of the \Hot population. The fact however, remains that the \Hot population has a flat size distribution over all observable sizes.

The behaviour of the \Hot population luminosity function at the faint end is uncertain; the shallow slope of $\alpha_{\Hot}\sim0.35$ for $R\lesssim25$ is compatible with the lack of faint detections of the \Hot population in fainter surveys, eliminating the need for a break at the faint end.  If however, the \Hot population slope is as steep as the upper limits of our confidence interval, a break at magnitudes fainter than $R\sim 26$ is still required. The existence of a break in the \Hot population will only be confirmed by an off-ecliptic survey sensitive to objects with $R>26$. 

The turn-up, as proposed by \citet{Morbidelli2009} would produce an abundance of faint objects in our observed \Hot population. We can test this utilizing the Anderson-Darling statistic. We generate a fake \Hot subsample from a shallow power-law with the observed \Hot population slope that turns up to the best-fit observed slope of the \Cold population, $\alpha_{\Cold}=0.8$, at some magnitude $m_{TU}$. We then fit a power-law to that sample, and calculate the Anderson-Darling statistic. This is repeated, and the probability of finding a statistic worse than the observations is calculated versus the value of $m_{TU}$. The strength of this test is weakened by the sparse sampling of high-latitude sources in our dataset. We still however, eliminate such a turn-up to the slope of the \Cold population for $m_{TU}<24.2$ (r') at the 2$-\sigma$ level and $<23.6$ at the 3-$\sigma$ level. This is incompatible with the assertion of \citet{Morbidelli2009} who suggest that the \Hot population should have the same slope as the \Cold population over an absolute magnitude range of $6.5<H<9$, corresponding to r' magnitudes, $22\lesssim r'\lesssim 24.5$. The significance of the test results over the entire magnitude range however, prevents this result from being iron clad. Clearly, the possibility of a turn-up (or down) must be tested from a survey which detects a large sample of high-latitude objects.

The difference in size distribution slopes of the \Hot and \Cold populations suggest very different histories for these two groups. Interestingly, the shallow slope of the \Hot distribution is compatible with a heavily collisionally processed population, either one that has reached collisional equilibrium, or in which the largest objects are fragments of even larger, disrupted primordial bodies \citep{Obrien2003,Bottke2010}. For collisional evolution to completely reshape the size distribution of the \Hot population would require extremely high collision rates, seemingly incompatible with plausible protoplanetary disk densities, and formation scenarios \citep{Stern1996a,Kenyon2004,Fraser2009c}. Rather it is likely that this slope is the result of the accretionary processes that formed the \Hot population.

Unlike the \Hot population, the steep slope of the \Cold population is entirely incompatible with a collisionally evolved distribution. This slope must be the result of accretionary processes as well. 

The shallower slope of the \Hot population, and the fact that the largest objects of the \Hot population are larger than those of the \Cold population, implies that the \Hot population achieved a more advanced stage of accretion than the \Cold population \citep{Kenyon2002}. The incompatibility of their slopes implies that these two populations underwent different accretion scenarios. It is possible, that the \Hot population underwent a longer duration of accretion than the \Cold population. To halt accretion for the \Cold population would require excitation and mass depletion with some yet unseen perturber beyond the outer edge of the Kuiper belt. The past existence of one or more planetary embryos in the Kuiper belt region is a possible source of this excitation, and  has been proposed to explain some of the Kuiper belt dynamical features \citep{Gladman2006,Lykawka2008}. 

Another possibility is that the \Hot population underwent accretion in a more dense region of the protoplanetary nebula. This would result in more rapid accretion than compared to the \Cold population, allowing the \Hot population to grow to larger sizes before the process was halted. Under this scenario, the \Hot and \Cold populations could then have similar accretion timescales. As currently favoured formation scenarios suggest the \Hot population was scattered from a region closer to the Sun than their current locations \citep{Malhotra1993,Gomes2003,Levison2008}, where the protoplanetary disk might have been more dense, this scenario seems likely.

However the \Hot and \Cold populations came about, their incompatible size distributions imply different accretion scenarios. Once accretion was finished, these objects were excited and emplaced onto their current orbits, creating the architecture of the current Kuiper belt. Given the dynamically cold nature of the \Cold population, it seems plausible that this population formed in-situ. What ever the mechanism(s) ultimately responsible for the belt, our results show that little mixing between the \Hot and \Cold populations has occurred.

\subsection{Comparison with Other Populations}
There is a striking similarity in the size distributions of the \Cold population, and the Jupiter Trojans. Both have similarly steep slopes, ($q_{\Cold}\sim5.1$ and $q_{JT}\sim 5.5$), and both exhibit breaks to shallower slopes at roughly the same object diameter \citep{Jewitt2000,Fraser2009}. This result argues strongly against the Trojan formation mechanism of the so-called NICE model in which the Trojan populations and the hot and cold Kuiper belt populations are all scattered into their current regions from the same primordial disk population \citep{Morbidelli2005,Nesvorny2009}. If this scenario were true, \citet{Morbidelli2009} has shown that over the magnitude range of our observations, the hot population should have the same slope as that observed for large Jupiter trojans. Using the Anderson-Darling statistic, we tested the probability of the \Hot sample being drawn from a luminosity function with the slope of the Jupiter trojan luminosity function, $\alpha_{JT}=0.9$. We found that the \Hot sample could not be drawn from the Jupiter trojan population at greater than the 3-$\sigma$ significance. As the size distributions of these two populations in the size range considered have not evolved significantly since they were emplaced in their current regions \citep{Davis2002,Fraser2009}, we must conclude that the hot and Jupiter trojan populations must have different progenator populations. This result and the lack of mixing between the hot and cold populations is difficult to reconcile with the NICE model. It seems likely the hot, cold, and trojan populations formed by separate means. If this is true, the similarities between the cold and trojan populations are quite a coincidence.

Another similarity is seen between the \Hot population ($q_{\Hot}\sim 3$) and the Jupiter family comets, who exhibit a size distribution slope of $q_{JFC}\sim2.8$ albeit over a smaller size range, $1\lesssim D \lesssim 10$ km \citep{Tancredi2006,WeissmanDPS2009}. This resemblance is interesting as one likely source of Jupiter family comets are scattered disk objects - members of the \Hot population -  which have fallen into the inner Solar system under gravitational perturbations from the gas-giants, suggesting that the scattered disk size distribution might be a power-law for $D\gtrsim1$ km. We consider this possibility here.

While the observations we present do not measure the size distribution of the scattered disk directly, the similarity in slopes between the \Hot and \Close subsamples implies that the scattered disk cannot have a size distribution significantly different than that of the \Hot sample as a whole. Additionally we cannot determine the total number scattered disk objects from our observations. Rather, we turn to other estimates which suggest that, if their size distribution is a power-law with slope of $q\sim 3$, then there are roughly $10^7-10^8$ scattered disk objects with $1\lesssim D \lesssim10$ km \citep[][Schwamb, personal communication]{Trujillo2000,Parker2010b}. 

Simulations by \citet{Volk2008} suggest that if the scattered disk is the sole source of the Jupiter family comets, then there must be at least $10^9$ objects in that population to account for the current flux of Jupiter family comets through the inner Solar system, implying that either the scattered disk is not the sole source of the Jupiter family comets, or that the extrapolation of a power-law to $D\sim 1$ km is unreasonable; it must be that the size distribution of scattered disk objects steepens significantly in the $D\sim10-100$ km range. Further observations are required before this feature can be detected.

\section{Conclusions} \label{sec:conclusions}
We have performed an ecliptic survey, and have detected 88 Kuiper belt objects with magnitudes $21<r'_{Sloan}<25.2$. A subset of these objects have received additional follow-up observations allowing us to accurately determine their inclinations and distances. Using these data, we have measured the size distribution of the \Hot and \Cold subsamples, historically defined as those objects with inclinations above and below $5^o$ respectively. This measurement, which is independent of any previous observations has confirmed that the \Cold population has a much steeper luminosity function, with slope $\alpha_{Cold}=0.82\pm0.23$, than the \Hot population, with slope $\alpha_{\Hot}=0.35\pm0.21$.

The observed luminosity functions imply different size distributions for the \Hot and \Cold populations. The size distribution slopes of the the two subsamples are $q_{\Cold}=5.1\pm1.1$ and $q_{\Hot}=2.8\pm1.0$.

In addition, we have found that the \Close population, which is defined as those objects with heliocentric distance, $d<38$ have the same luminosity function slope as the \Hot  subsample, demonstrating that these two share a similar size distribution. In addition, our findings suggest that the dynamical populations which make up both the \Hot and \Close populations must all have similar size distributions. 

The primary consequence of these findings is that the \Cold population, which consists primarily of cold classical Kuiper belt objects, has a separate, and distinct accretion history from the \Hot population, requiring either potentially different accretion timescales for the two populations, or formation in different locations of the protoplanetary disk. These observations reveal the similarities in the size distributions between the subsamples of the Kuiper belt and the Jupiter family comets and Trojans, suggesting a connection between the formation and subsequent evolution of the small body populations of the outer Solar system.

\section{Acknowledgements}

Based in part on data collected at Subaru Telescope, which is operated by the National Astronomical Observatory of Japan. Some of the data presented herein were obtained at the W. M. Keck Observatory, which is operated as a scientific partnership among the California Institute of Technology, the University of California, and the National Aeronautics and Space Administration.  This research used the facilities of the Canadian Astronomy Data Centre operated by the National Research Council of Canada with the support of the Canadian Space Agency. The research upon which this paper is based was supported by National Aeronautic and Space Administration (NASA) Grant No. NASA.000261

%% References with bibTeX database:
%\bibliographystyle{elsarticle-harv}
%\bibliography{AstroElsart}

%%%Examples incase you need them
\newpage
\onecolumn

\begin{longtable}{ll}

Right Ascension & Declination \\ \hline
01:45:59.0 & +10:03:00.0 \\ 
01:46:29.7 & +11:24:00.0 \\ 
01:48:27.6 & +10:30:00.0 \\ 
01:48:59.1 & +11:51:00.0 \\ 
01:50:46.1 & +10:30:00.0 \\ 
01:51:07.5 & +11:24:00.0 \\ 
01:51:29.0 & +12:18:00.0 \\ 
01:51:39.8 & +12:45:00.0 \\ 
01:52:53.7 & +10:03:00.0 \\ 
01:53:15.5 & +10:57:00.0 \\ 
01:53:37.3 & +11:51:00.0 \\ 
01:53:48.3 & +12:18:00.0 \\ 
01:54:10.4 & +13:12:00.0 \\ 
01:55:34.1 & +10:57:00.0 \\ 
01:55:45.3 & +11:24:00.0 \\ 
01:56:30.2 & +13:12:00.0 \\ 
01:58:15.6 & +11:51:00.0 \\ 
01:58:50.0 & +13:12:00.0 \\ 
02:00:34.7 & +11:51:00.0 \\ 
02:07:32.1 & +11:51:00.0 \\ 
02:07:44.5 & +12:18:00.0 \\ 
02:07:56.8 & +12:45:00.0 \\ 
02:08:09.3 & +13:12:00.0 \\ 
02:09:51.3 & +11:51:00.0 \\ 
02:10:16.4 & +12:45:00.0 \\ 
02:12:23.2 & +12:18:00.0 \\ 
02:13:01.8 & +13:39:00.0 \\ 
02:15:08.7 & +13:12:00.0 \\ 
02:15:21.8 & +13:39:00.0 \\ 
02:17:15.2 & +12:45:00.0 \\ 
02:17:28.5 & +13:12:00.0 \\ 
02:21:54.3 & +12:45:00.0 \\ 
02:22:21.9 & +13:39:00.0 \\ 
02:22:35.8 & +14:06:00.0 \\ 
02:24:13.9 & +12:45:00.0 \\ 
\hline
\caption{Suprime-cam Discovery field centers. Coordinates are presented in the J2000 epoch. \label{tab:pointings}}
\end{longtable}

\clearpage

\begin{longtable}{lccccc}

	Object \footnotemark[1] & $r'_{Subaru}$ & $r'_{SDSS}$ & $(g_{SDSS}-r'_{SDSS})$ & R (AU) \footnotemark[2] & i ($^o$) \footnotemark[2] \\	
	\hline
obj\_0 & $25.2 \pm 0.1$ & $25.0 \pm 0.1$ & - & $42.0 \pm 3.0$ & $10.0 \pm 9.0$ \\
obj\_1 & $24.55 \pm 0.07$ & $24.52 \pm 0.07$ & - & $42.0 \pm 3.0$ & $2.0 \pm 8.0$ \\
obj\_2 & $24.75 \pm 0.08$ & $24.72 \pm 0.08$ & - & $40.0 \pm 3.0$ & $20.0 \pm 10.0$ \\
Nobj\_3 & $23.99 \pm 0.05$ & $24.02 \pm 0.05$ & $0.46 \pm 0.09$ & $41.0 \pm 3.0$ & $4.0 \pm 2.0$ \\
obj\_4 & $24.71 \pm 0.08$ & $24.65 \pm 0.08$ & - & $43.0 \pm 3.0$ & $1.0 \pm 6.0$ \\
Nobj\_5 & $24.44 \pm 0.07$ & $24.45 \pm 0.05$ & $0.72 \pm 0.09$ & $43.5 \pm 0.2$ & $0.7 \pm 0.03$ \\
obj\_6 & $23.5 \pm 0.04$ & $23.47 \pm 0.04$ & - & $35.0 \pm 3.0$ & $6.0 \pm 7.0$ \\
obj\_7 & $24.83 \pm 0.09$ & $24.78 \pm 0.09$ & - & $41.0 \pm 3.0$ & $15.0 \pm 10.0$ \\
Nobj\_8 & $23.74 \pm 0.04$ & $23.73 \pm 0.04$ & $0.4 \pm 0.1$ & $45.0 \pm 3.0$ & $1.67 \pm 0.01$ \\
obj\_9 & $24.33 \pm 0.07$ & $24.29 \pm 0.07$ & - & $35.0 \pm 2.0$ & $2.0 \pm 6.0$ \\
Nobj\_10 & $23.02 \pm 0.03$ & $22.98 \pm 0.03$ & $0.7 \pm 0.1$ & $42.0 \pm 3.0$ & $6.0 \pm 2.0$ \\
Nobj\_11 & $24.19 \pm 0.05$ & $24.15 \pm 0.05$ & $0.6 \pm 0.1$ & $39.6 \pm 0.4$ & $7.3 \pm 0.6$ \\
Nobj\_12 & $23.55 \pm 0.04$ & $23.55 \pm 0.04$ & $0.53 \pm 0.08$ & $41.0 \pm 3.0$ & $30.0 \pm 10.0$ \\
obj\_13 & $24.63 \pm 0.08$ & $24.55 \pm 0.07$ & - & $40.0 \pm 3.0$ & $5.0 \pm 8.0$ \\
obj\_14 & $24.1 \pm 0.05$ & $24.07 \pm 0.05$ & - & $32.0 \pm 2.0$ & $3.0 \pm 5.0$ \\
obj\_15 & $24.9 \pm 0.09$ & $24.88 \pm 0.09$ & - & $30.0 \pm 3.0$ & $30.0 \pm 20.0$ \\
obj\_16 & $25.0 \pm 0.1$ & $24.96 \pm 0.1$ & - & $43.0 \pm 3.0$ & $1.0 \pm 8.0$ \\
Nobj\_17 & $21.65 \pm 0.02$ & $21.63 \pm 0.02$ & $0.6 \pm 0.09$ & $38.5 \pm 0.1$ & $9.6 \pm 0.2$ \\
Nobj\_18 & $24.43 \pm 0.06$ & $24.25 \pm 0.06$ & $0.9 \pm 0.1$ & $45.0 \pm 2.0$ & $0.97 \pm 0.01$ \\
obj\_19 & $24.88 \pm 0.09$ & $24.86 \pm 0.09$ & - & $36.0 \pm 3.0$ & $9.0 \pm 7.0$ \\
Nobj\_20 & $24.01 \pm 0.05$ & $24.0 \pm 0.05$ & $0.45 \pm 0.09$ & $44.0 \pm 3.0$ & $0.7 \pm 0.2$ \\
Nobj\_21 & $23.82 \pm 0.04$ & $23.79 \pm 0.04$ & $0.4 \pm 0.09$ & $39.0 \pm 5.0$ & $40.0 \pm 30.0$ \\
obj\_22 & $24.65 \pm 0.08$ & $24.62 \pm 0.08$ & - & $46.0 \pm 3.0$ & $10.0 \pm 10.0$ \\
Nobj\_23 & $23.96 \pm 0.05$ & $23.96 \pm 0.05$ & $0.5 \pm 0.1$ & $37.0 \pm 2.0$ & $1.7 \pm 0.6$ \\
obj\_24 & $24.75 \pm 0.08$ & $24.7 \pm 0.08$ & - & $42.0 \pm 3.0$ & $3.0 \pm 8.0$ \\
obj\_25 & $24.86 \pm 0.09$ & $24.82 \pm 0.09$ & - & $43.0 \pm 3.0$ & $3.0 \pm 8.0$ \\
obj\_26 & $25.0 \pm 0.1$ & $25.0 \pm 0.1$ & - & $43.0 \pm 3.0$ & $2.0 \pm 8.0$ \\
Nobj\_27 & $24.33 \pm 0.06$ & $24.38 \pm 0.05$ & $0.33 \pm 0.07$ & $38.0 \pm 3.0$ & $23.0 \pm 9.0$ \\
2004 VT75 \footnotemark[3] & $21.92 \pm 0.03$ & $21.9 \pm 0.03$ & - & 37.55 & 12.823 \\
obj\_29 & $23.65 \pm 0.04$ & $23.62 \pm 0.04$ & - & $30.0 \pm 3.0$ & $19.0 \pm 9.0$ \\
Nobj\_30 & $24.36 \pm 0.06$ & $24.29 \pm 0.06$ & $0.6 \pm 0.1$ & $46.0 \pm 3.0$ & $2.2 \pm 0.8$ \\
Nobj\_31 & $23.86 \pm 0.04$ & $23.78 \pm 0.04$ & $0.69 \pm 0.09$ & $43.0 \pm 2.0$ & $4.0 \pm 1.0$ \\
Nobj\_32 & $22.07 \pm 0.02$ & $22.08 \pm 0.02$ & $0.36 \pm 0.09$ & $39.39 \pm 0.01$ & $9.45 \pm 0.01$ \\
Nobj\_33 & $23.83 \pm 0.04$ & $23.79 \pm 0.04$ & $0.7 \pm 0.1$ & $38.0 \pm 2.0$ & $5.0 \pm 2.0$ \\
Nobj\_34 & $23.53 \pm 0.05$ & $23.54 \pm 0.04$ & $0.6 \pm 0.1$ & $37.0 \pm 2.0$ & $10.0 \pm 4.0$ \\
Nobj\_35 & $22.9 \pm 0.03$ & $22.88 \pm 0.03$ & $0.47 \pm 0.08$ & $36.2 \pm 0.1$ & $20.9 \pm 0.4$ \\
obj\_36 & $21.57 \pm 0.02$ & $21.55 \pm 0.02$ & - & $36.0 \pm 3.0$ & $14.0 \pm 8.0$ \\
obj\_37 & $22.51 \pm 0.02$ & $22.49 \pm 0.03$ & - & $32.0 \pm 3.0$ & $20.0 \pm 10.0$ \\
obj\_38 & $24.52 \pm 0.07$ & $24.49 \pm 0.07$ & - & $32.0 \pm 2.0$ & $11.0 \pm 7.0$ \\
Nobj\_39 & $23.9 \pm 0.05$ & $23.91 \pm 0.04$ & $0.58 \pm 0.09$ & $43.0 \pm 3.0$ & $2.9 \pm 0.9$ \\
Nobj\_40 & $23.25 \pm 0.03$ & $23.21 \pm 0.03$ & $0.66 \pm 0.09$ & $41.0 \pm 2.0$ & $2.6 \pm 0.9$ \\
Nobj\_41 & $23.77 \pm 0.05$ & $23.74 \pm 0.05$ & $0.66 \pm 0.09$ & $43.0 \pm 3.0$ & $2.6 \pm 1.0$ \\
obj\_42 & $24.26 \pm 0.06$ & $24.23 \pm 0.06$ & - & $32.0 \pm 3.0$ & $30.0 \pm 10.0$ \\
obj\_43 & $24.74 \pm 0.08$ & $24.72 \pm 0.08$ & - & $46.0 \pm 3.0$ & $8.0 \pm 10.0$ \\
Nobj\_44 & $24.18 \pm 0.05$ & $24.15 \pm 0.05$ & $0.38 \pm 0.09$ & $46.0 \pm 3.0$ & $2.8 \pm 1.0$ \\
obj\_45 & $24.55 \pm 0.07$ & $24.42 \pm 0.07$ & - & $47.0 \pm 3.0$ & $1.0 \pm 10.0$ \\
Nobj\_46 & $23.83 \pm 0.04$ & $23.8 \pm 0.04$ & $0.5 \pm 0.1$ & $37.0 \pm 3.0$ & $18.0 \pm 7.0$ \\
Nobj\_47 & $24.31 \pm 0.06$ & $24.27 \pm 0.05$ & $0.57 \pm 0.09$ & $43.0 \pm 0.2$ & $0.871 \pm 0.002$ \\
obj\_48 & $25.3 \pm 0.2$ & $25.1 \pm 0.1$ & - & $19.0 \pm 4.0$ & $40.0 \pm 30.0$ \\
Nobj\_49 & $23.85 \pm 0.04$ & $23.78 \pm 0.04$ & $0.56 \pm 0.09$ & $44.0 \pm 3.0$ & $7.0 \pm 2.0$ \\
Nobj\_50 & $23.57 \pm 0.04$ & $23.53 \pm 0.04$ & $0.65 \pm 0.09$ & $45.0 \pm 2.0$ & $4.0 \pm 1.0$ \\
obj\_51 & $25.0 \pm 0.1$ & $24.95 \pm 0.1$ & - & $34.0 \pm 2.0$ & $5.0 \pm 6.0$ \\
obj\_52 & $24.47 \pm 0.07$ & $24.41 \pm 0.07$ & - & $35.0 \pm 3.0$ & $4.0 \pm 6.0$ \\
Nobj\_53 & $23.92 \pm 0.06$ & $23.81 \pm 0.06$ & $0.7 \pm 0.1$ & $46.0 \pm 3.0$ & $2.2 \pm 0.6$ \\
obj\_54 & $24.95 \pm 0.1$ & $24.78 \pm 0.09$ & - & $43.0 \pm 3.0$ & $3.0 \pm 8.0$ \\
obj\_55 & $24.98 \pm 0.1$ & $24.93 \pm 0.1$ & - & $42.0 \pm 3.0$ & $2.0 \pm 8.0$ \\
Nobj\_56 & $24.35 \pm 0.06$ & $24.27 \pm 0.06$ & $0.6 \pm 0.1$ & $47.0 \pm 3.0$ & $3.0 \pm 1.0$ \\
obj\_57 & $24.84 \pm 0.09$ & $24.81 \pm 0.09$ & - & $43.0 \pm 3.0$ & $20.0 \pm 10.0$ \\
obj\_58 & $25.1 \pm 0.1$ & $24.9 \pm 0.1$ & - & $24.0 \pm 3.0$ & $30.0 \pm 10.0$ \\
Nobj\_59 & $24.44 \pm 0.07$ & $24.22 \pm 0.05$ & $0.66 \pm 0.07$ & $41.0 \pm 0.1$ & $2.14 \pm 0.05$ \\
obj\_60 & $24.05 \pm 0.05$ & $24.01 \pm 0.05$ & - & $28.0 \pm 2.0$ & $6.0 \pm 5.0$ \\
Nobj\_61 & $23.64 \pm 0.05$ & $23.57 \pm 0.05$ & $0.65 \pm 0.09$ & $42.0 \pm 2.0$ & $2.9 \pm 0.9$ \\
Nobj\_62 & $24.16 \pm 0.07$ & $24.21 \pm 0.06$ & $0.44 \pm 0.09$ & $43.0 \pm 3.0$ & $13.0 \pm 5.0$ \\
obj\_63 & $23.68 \pm 0.04$ & $23.65 \pm 0.04$ & - & $22.0 \pm 2.0$ & $14.0 \pm 7.0$ \\
obj\_64 & $24.54 \pm 0.07$ & $24.52 \pm 0.07$ & - & $42.0 \pm 3.0$ & $2.0 \pm 7.0$ \\
Nobj\_65 & $23.3 \pm 0.03$ & $23.27 \pm 0.03$ & $0.6 \pm 0.1$ & $43.5 \pm 0.1$ & $0.82 \pm 0.01$ \\
obj\_66 & $24.29 \pm 0.06$ & $24.26 \pm 0.06$ & - & $20.0 \pm 8.0$ & $70.0 \pm 80.0$ \\
Nobj\_67 & $24.44 \pm 0.06$ & $24.36 \pm 0.06$ & $0.67 \pm 0.09$ & $45.0 \pm 3.0$ & $1.6 \pm 0.3$ \\
Nobj\_68 & $23.81 \pm 0.04$ & $23.74 \pm 0.04$ & $0.64 \pm 0.09$ & $41.0 \pm 3.0$ & $3.0 \pm 1.0$ \\
obj\_69 & $24.65 \pm 0.08$ & $24.57 \pm 0.08$ & - & $42.0 \pm 3.0$ & $3.0 \pm 8.0$ \\
obj\_70 & $24.53 \pm 0.07$ & $24.5 \pm 0.07$ & - & $41.0 \pm 3.0$ & $4.0 \pm 8.0$ \\
obj\_71 & $23.85 \pm 0.04$ & $23.82 \pm 0.05$ & - & $34.0 \pm 2.0$ & $3.0 \pm 5.0$ \\
Nobj\_72 & $23.91 \pm 0.05$ & $23.88 \pm 0.05$ & $0.6 \pm 0.1$ & $44.0 \pm 2.0$ & $1.4 \pm 0.3$ \\
Nobj\_73 & $23.94 \pm 0.05$ & $23.83 \pm 0.05$ & $0.6 \pm 0.1$ & $44.0 \pm 2.0$ & $2.1 \pm 0.8$ \\
Nobj\_74 & $24.41 \pm 0.08$ & $24.68 \pm 0.06$ & $0.35 \pm 0.08$ & $41.0 \pm 2.0$ & $4.0 \pm 1.0$ \\
obj\_75 & $25.2 \pm 0.1$ & $25.1 \pm 0.1$ & - & $36.0 \pm 3.0$ & $1.0 \pm 5.0$ \\
obj\_76 & $24.68 \pm 0.08$ & $24.6 \pm 0.08$ & - & $42.0 \pm 3.0$ & $2.0 \pm 8.0$ \\
Nobj\_77 & $22.74 \pm 0.03$ & $22.72 \pm 0.03$ & $0.6 \pm 0.1$ & $42.3 \pm 0.1$ & $1.24 \pm 0.02$ \\
Nobj\_78 & $23.74 \pm 0.04$ & $23.71 \pm 0.04$ & $0.67 \pm 0.09$ & $42.0 \pm 3.0$ & $2.1 \pm 0.3$ \\
obj\_79 & $24.85 \pm 0.09$ & $24.78 \pm 0.09$ & - & $50.0 \pm 3.0$ & $1.0 \pm 6.0$ \\
obj\_80 & $24.58 \pm 0.07$ & $24.54 \pm 0.07$ & - & $42.0 \pm 3.0$ & $5.0 \pm 8.0$ \\
obj\_81 & $24.03 \pm 0.05$ & $24.01 \pm 0.05$ & - & $33.0 \pm 2.0$ & $6.0 \pm 6.0$ \\
Nobj\_82 & $24.19 \pm 0.05$ & $24.13 \pm 0.05$ & $0.78 \pm 0.09$ & $43.1 \pm 0.1$ & $0.31 \pm 0.01$ \\
obj\_83 & $23.06 \pm 0.03$ & $22.95 \pm 0.03$ & - & $35.0 \pm 3.0$ & $30.0 \pm 10.0$ \\
obj\_84 & $24.98 \pm 0.1$ & $24.94 \pm 0.1$ & - & $41.0 \pm 3.0$ & $2.0 \pm 8.0$ \\
obj\_85 & $24.54 \pm 0.07$ & $24.52 \pm 0.07$ & - & $30.0 \pm 3.0$ & $17.0 \pm 9.0$ \\
Nobj\_86 & $23.87 \pm 0.04$ & $23.84 \pm 0.04$ & $0.68 \pm 0.09$ & $42.0 \pm 3.0$ & $2.1 \pm 0.6$ \\
obj\_87 & $25.1 \pm 0.1$ & $25.1 \pm 0.1$ & - & $45.0 \pm 3.0$ & $8.0 \pm 9.0$ \\ 
\hline
	\caption{Discovery Subaru magnitudes and Sloan magnitudes.\label{tab:objects}}
	\footnotetext[1]{Objects prefaced with N and D have follow-up observations in November and December respectively.}
	\footnotetext[2]{Barycentric Distance and ecliptic inclination determined from \emph{fit\_radec} \citep{Bernstein2000}.}
	\footnotetext[3]{Object distance and inclination taken from the Minor Planet Center.}
\end{longtable}

\begin{table}
\begin{center}
\caption{Best-fit power-law parameters.\label{tab:best-fits}}
\begin{tabular}{lccc}
Sample & $\alpha$ & $m_o (r')$ & $P(\Delta>\Delta_{obs})$ \\ \hline
\All & $0.56\pm 0.1$ & $22.9 \pm 0.4$ & 0.64 \\
\ColdGood & $0.82 \pm 0.23$ & $23.8\pm0.3$ & 0.28\\
\Cold & $0.79 \pm 0.20$ & $23.8 \pm 0.3$ & 0.47\\
\HotGood & $0.35\pm 0.21$ & $24.3\pm0.7$ & 0.75\\
\Hot & $0.35 \pm 0.19$ & $24.3 \pm 0.7$ & 0.65\\
\Close & $0.40 \pm 0.15$ & $23.6 \pm 0.6$ & 0.68\\ \hline
\end{tabular}
\caption{The slopes of the \All, \Hot, \Cold, and \Close samples (rows 3, 5, and 6) are likely a few percent steeper than in actuality. This bias however, is much smaller than the uncertainties of those parameters (see Section~\ref{sec:methods}).}
\end{center}
\end{table}

\newpage

%%%put figure captions separately
\section*{Fig. Captions} 

\begin{figure}[h] %  figure placement: here, top, bottom, or page
   \centering
   \includegraphics[width=6in]{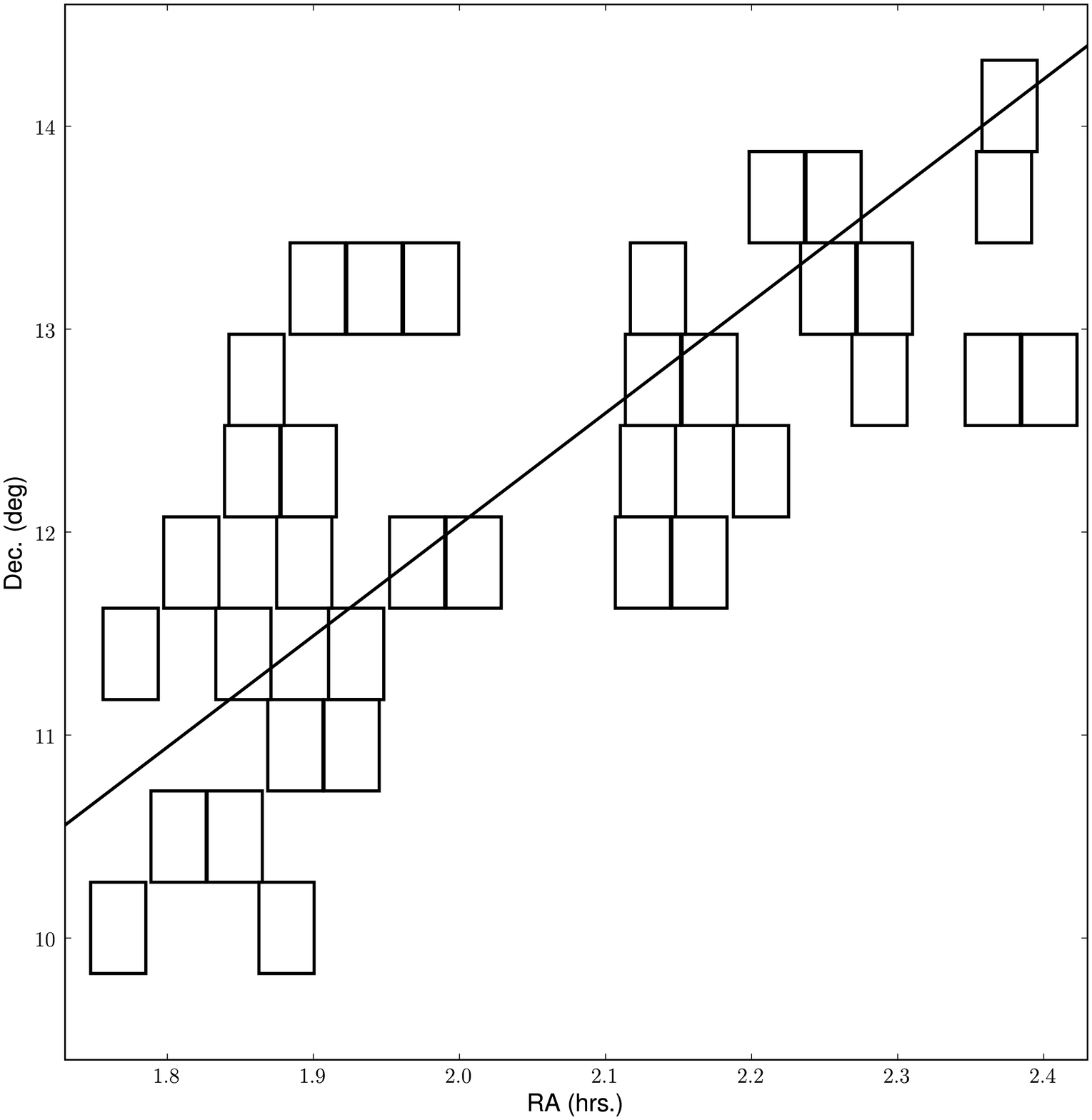}
   \caption{\label{fig:pointings}}
\end{figure}

Fig.~\ref{fig:pointings}. Diagram of the fields observed for moving objects during the Subaru observations. These fields were chosen to avoid bright stars. The field center coordinates are presented in Table~\ref{tab:pointings}.

\begin{figure}[h] %  figure placement: here, top, bottom, or page
   \centering
   \includegraphics[width=6in]{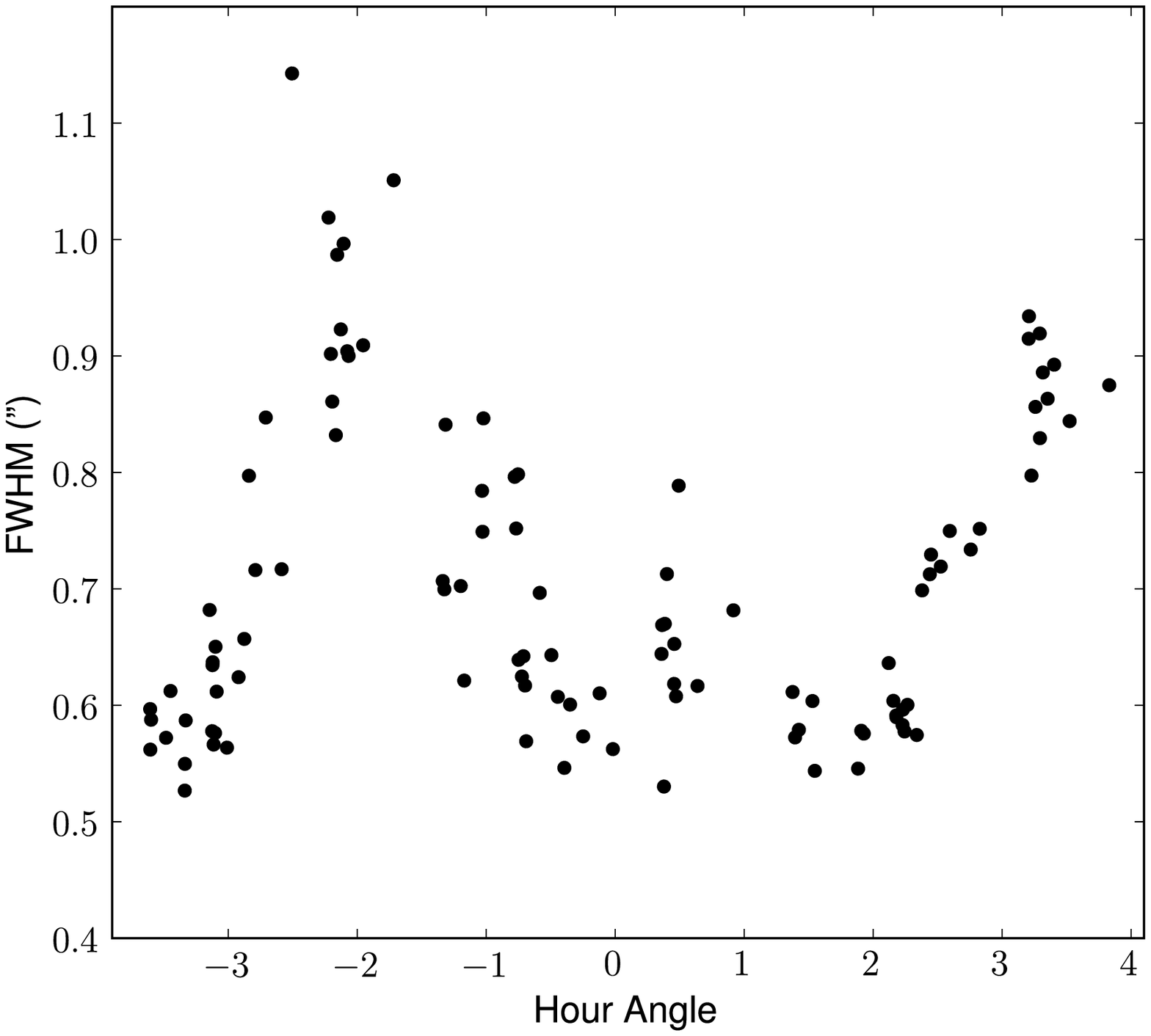}
   \caption{\label{fig:seeing}}
\end{figure}

Fig~\ref{fig:seeing}. Point source full-width at half maximum versus hour angle of all Subaru discovery observations.

\begin{figure}[h] %  figure placement: here, top, bottom, or page
   \centering
   \includegraphics[width=6in]{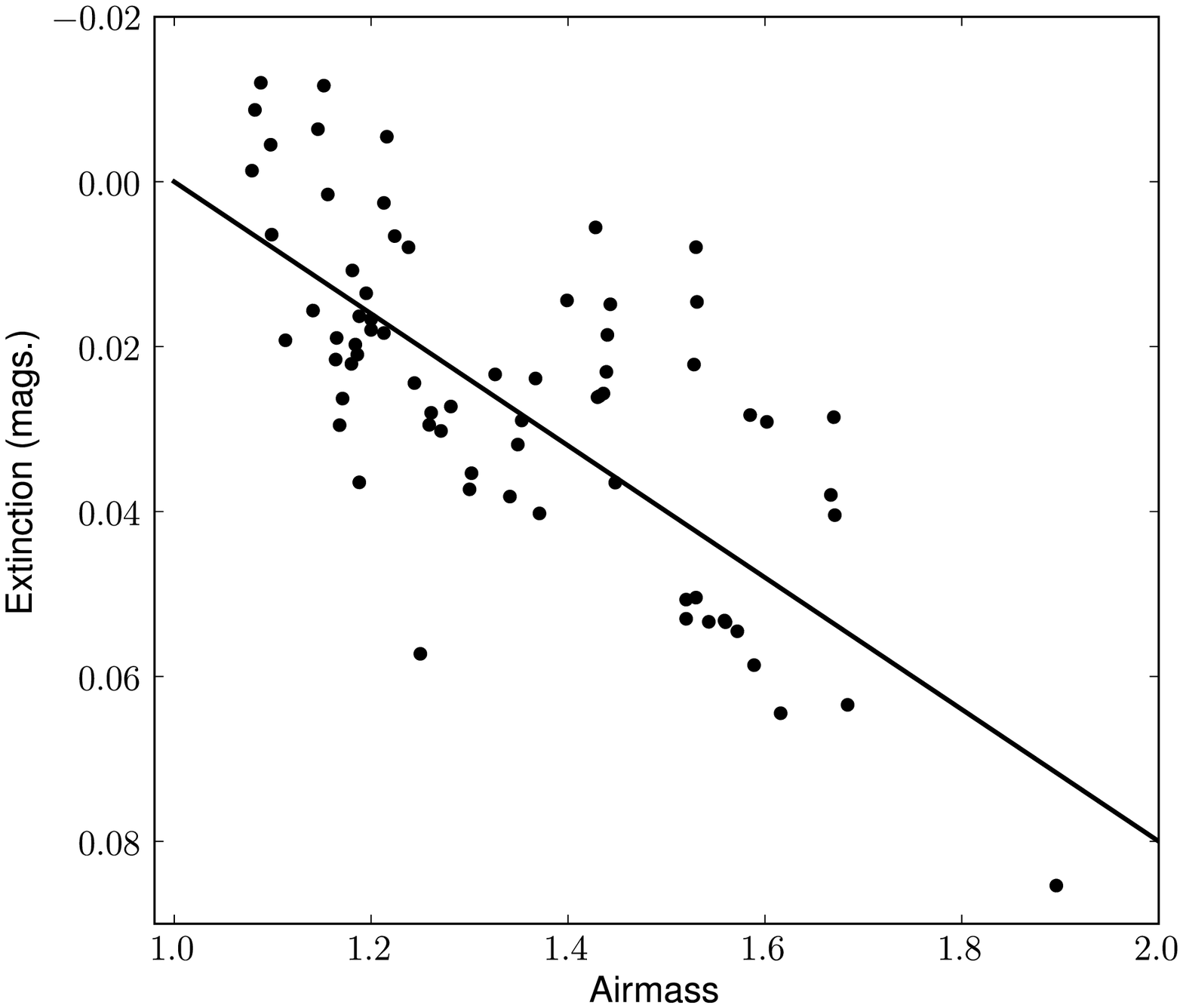}
   \caption{\label{fig:airmass}}
\end{figure}

Fig~\ref{fig:airmass}. Extinction versus airmass for the Subaru discovery observations. Each point is the average difference in magnitude of bright stars between common fields; each of the 35 discovery triplets has two data points. The line is the best-fit linear extinction law, with slope 0.08 magnitudes per unit airmass.  These data demonstrate the photometric nature of the discovery observations. 

\begin{figure}[h] %  figure placement: here, top, bottom, or page
   \centering
   \includegraphics[width=6in]{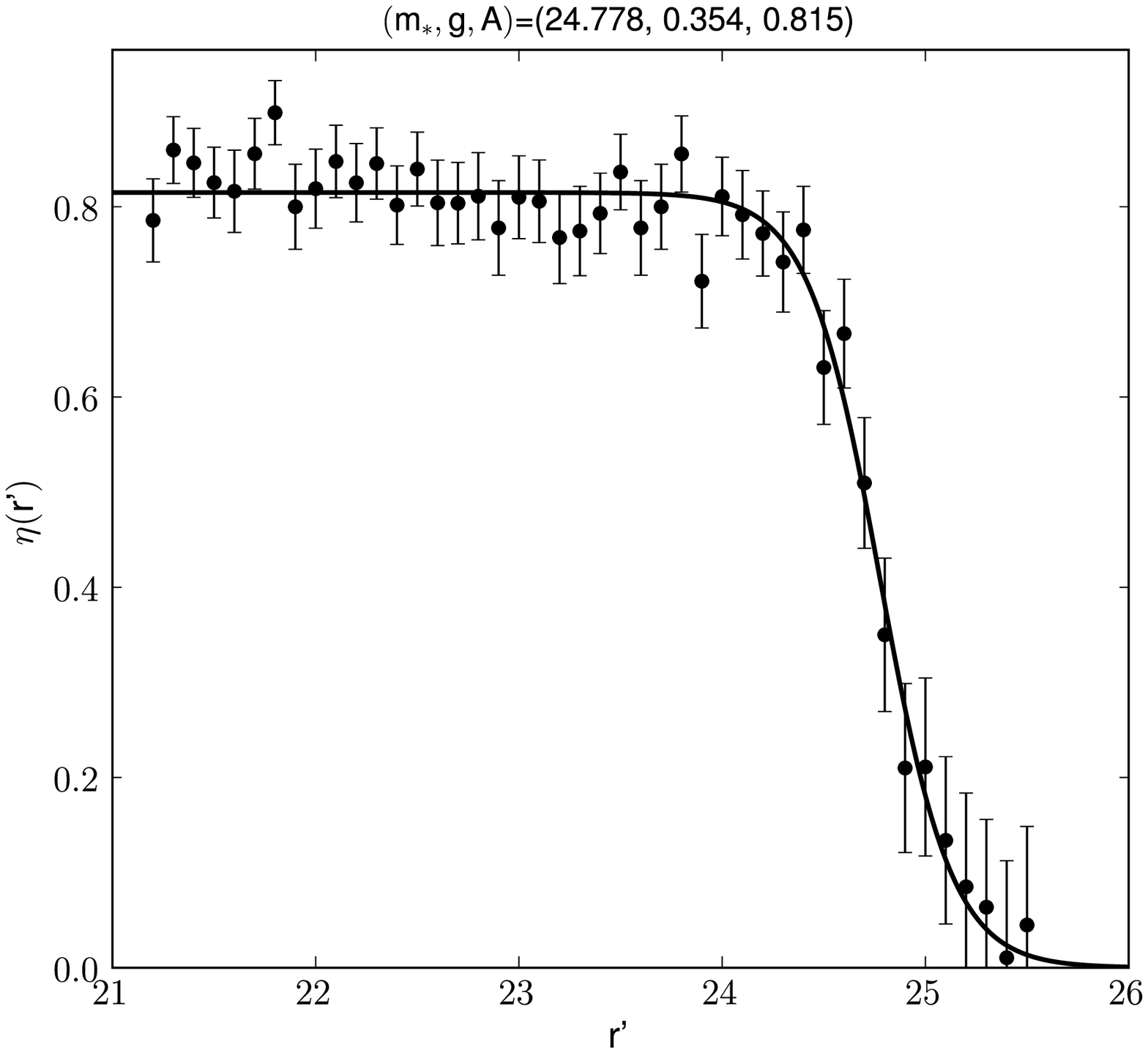}
   \caption{\label{fig:efficiency}}
\end{figure}

Fig~\ref{fig:efficiency}. Detection efficiency versus artificial source $r'_{Subaru}$ magnitude. Error-bars are 1-sigma poisson limits. Best-fit parameters of the efficiency function given by Equation~\ref{eq:efficiency} are presented at the top of the figure. 

\begin{figure}[h] %  figure placement: here, top, bottom, or page
   \centering
   \includegraphics[width=6in]{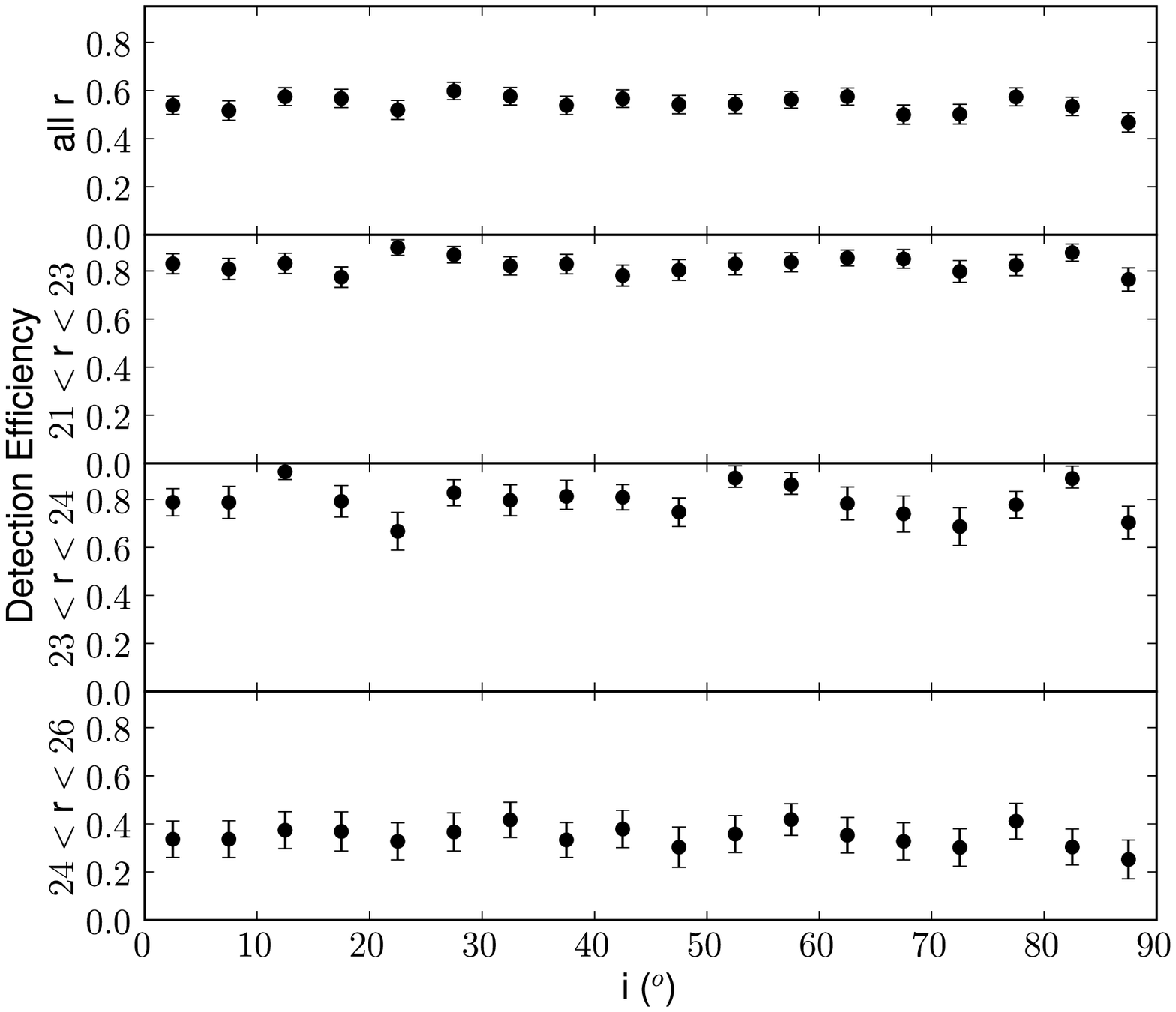}
   \caption{\label{fig:eff_i}}
\end{figure}

Fig~\ref{fig:eff_i}. Detection efficiency versus artificial object inclination. From top to bottom, all planted sources, and those with r'$<23$, $23<\mbox{r'}<24$, $24<\mbox{r'}<26$. This figure demonstrates that, given a particular object magnitude, our survey was equally sensitive to all inclinations.

\begin{figure}[h] %  figure placement: here, top, bottom, or page
   \centering
   \includegraphics[width=6in]{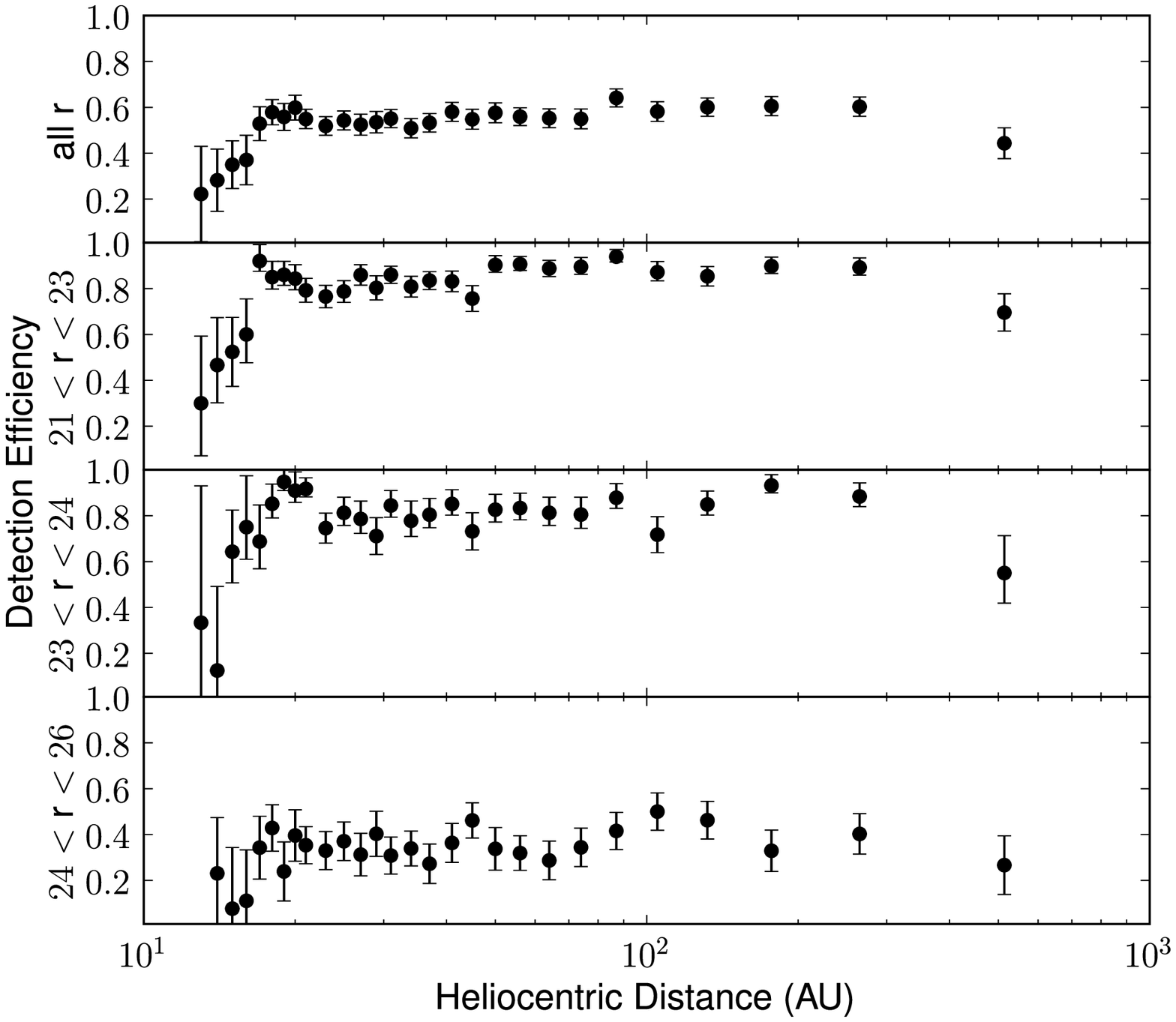}
   \caption{\label{fig:eff_r}}
\end{figure}

Fig~\ref{fig:eff_r}. Detection efficiency versus artificial object heliocentric distance. From top to bottom, all planted sources, and those with r'$<23$, $23<\mbox{r'}<24$, $24<\mbox{r'}<26$. This figure demonstrates that, given a particular object magnitude, no significant variation of detection efficiency with distance is apparent. Exceptions are at the extrema of the survey; our search was only sensitive to objects with apparent motions $0.2-10 "/\mbox{hr}$ roughly corresponding to distances $15\leq r \leq 800$ AU. 

\begin{figure}[h] %  figure placement: here, top, bottom, or page
   \centering
   \includegraphics[width=6in]{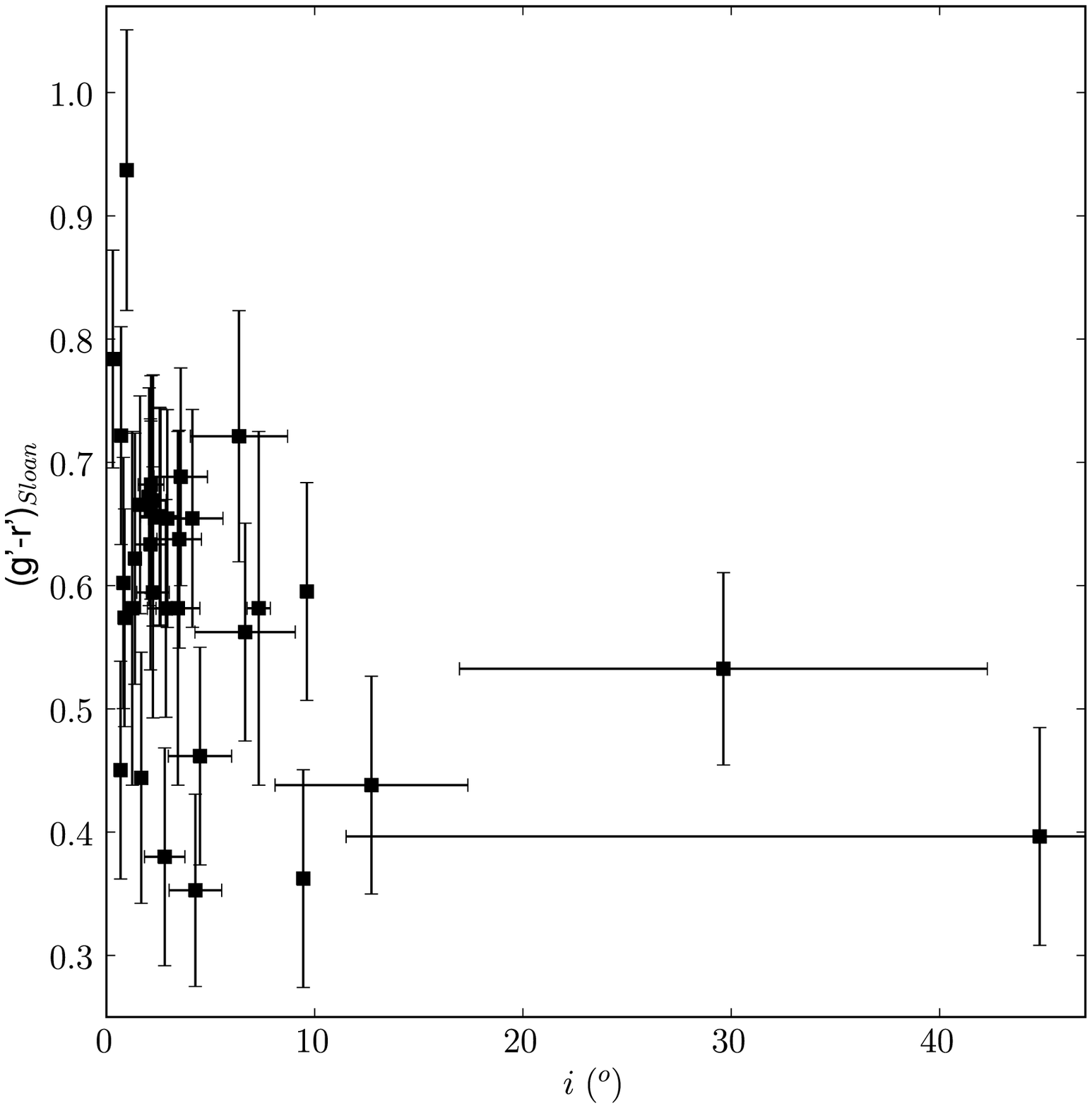}
   \caption{\label{fig:colour-inc}}
\end{figure}

Fig~\ref{fig:colour-inc}. Sloan colour versus inclination for those objects which received follow-up observations.

\begin{figure}[h] %  figure placement: here, top, bottom, or page
   \centering
   \includegraphics[width=6in]{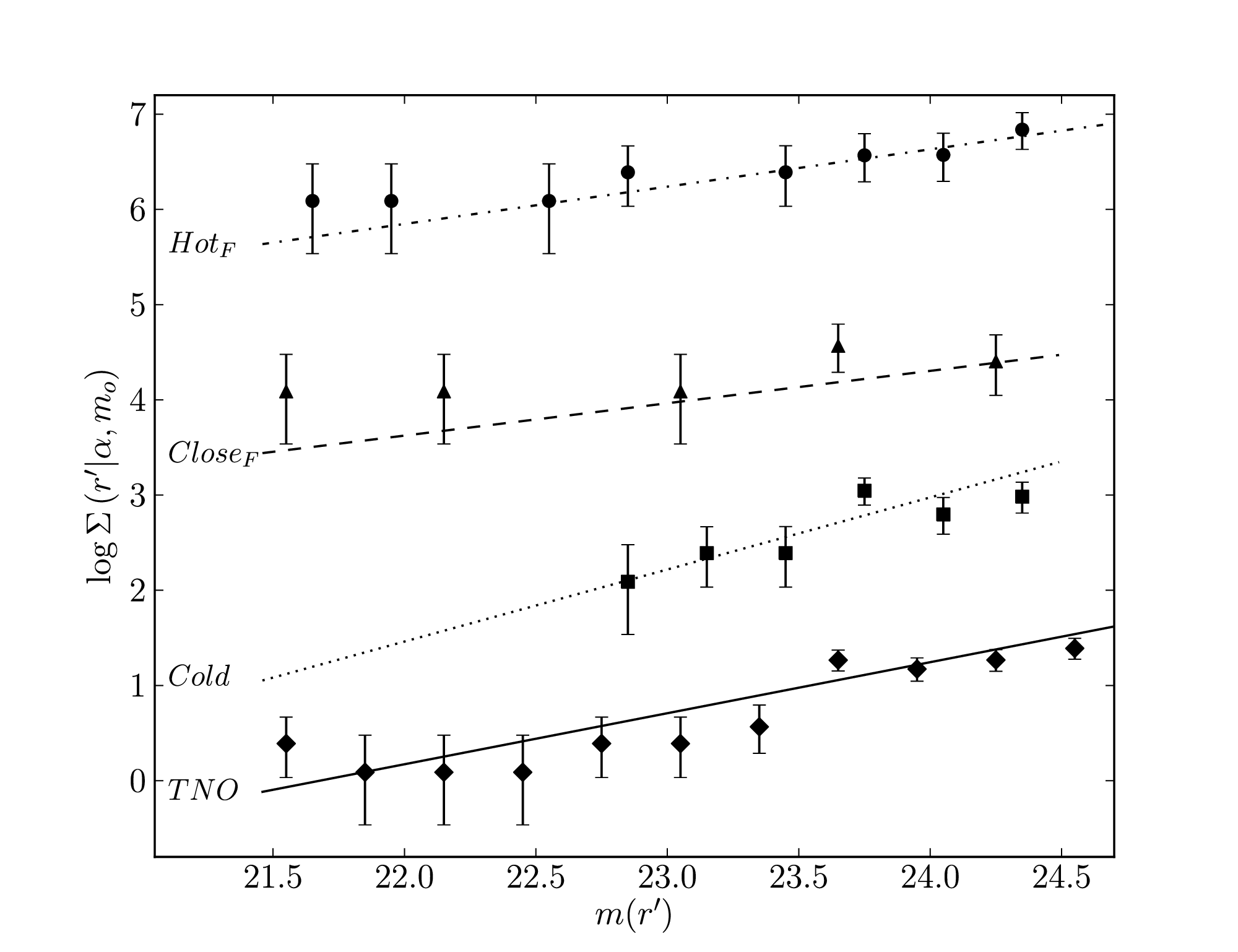}
   \caption{\label{fig:LF}}
\end{figure}

Fig~\ref{fig:LF}. The differential luminosity functions of the various samples defined in Section~\ref{sec:subsamples} offset in units of 2 for clarity. Diamonds: \All. Squares: \ColdGood. Triangles: \HotGood. Circles: \Close. Error-bars are the 1-$\sigma$ poisson limits for the number of objects in each bin. Lines are the best-fit power-laws, with values given in Table~\ref{tab:best-fits}.

\begin{figure}[h] %  figure placement: here, top, bottom, or page
   \centering
   \includegraphics[width=4in]{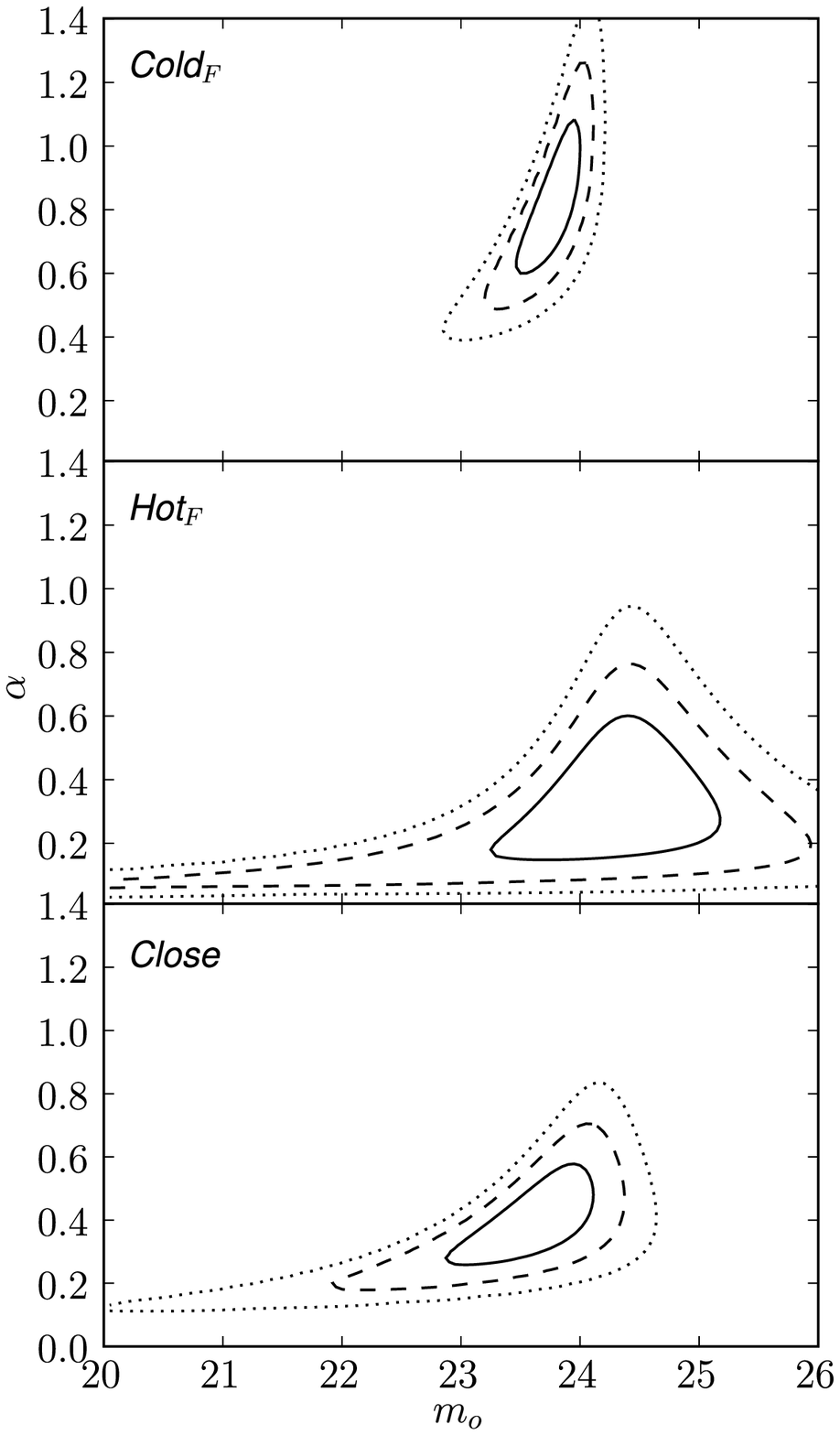}
   \caption{\label{fig:contours}}
\end{figure}

Fig~\ref{fig:contours}. The 1 (solid), 2 (dashed) and 3-$\sigma$ (dotted) likelihood contours of the fits presented in Figure~\ref{fig:LF} and Table~\ref{tab:best-fits}. The corresponding subsample is labeled in the top right of each panel.

\begin{figure}[h] %  figure placement: here, top, bottom, or page
   \centering
   \includegraphics[width=6in]{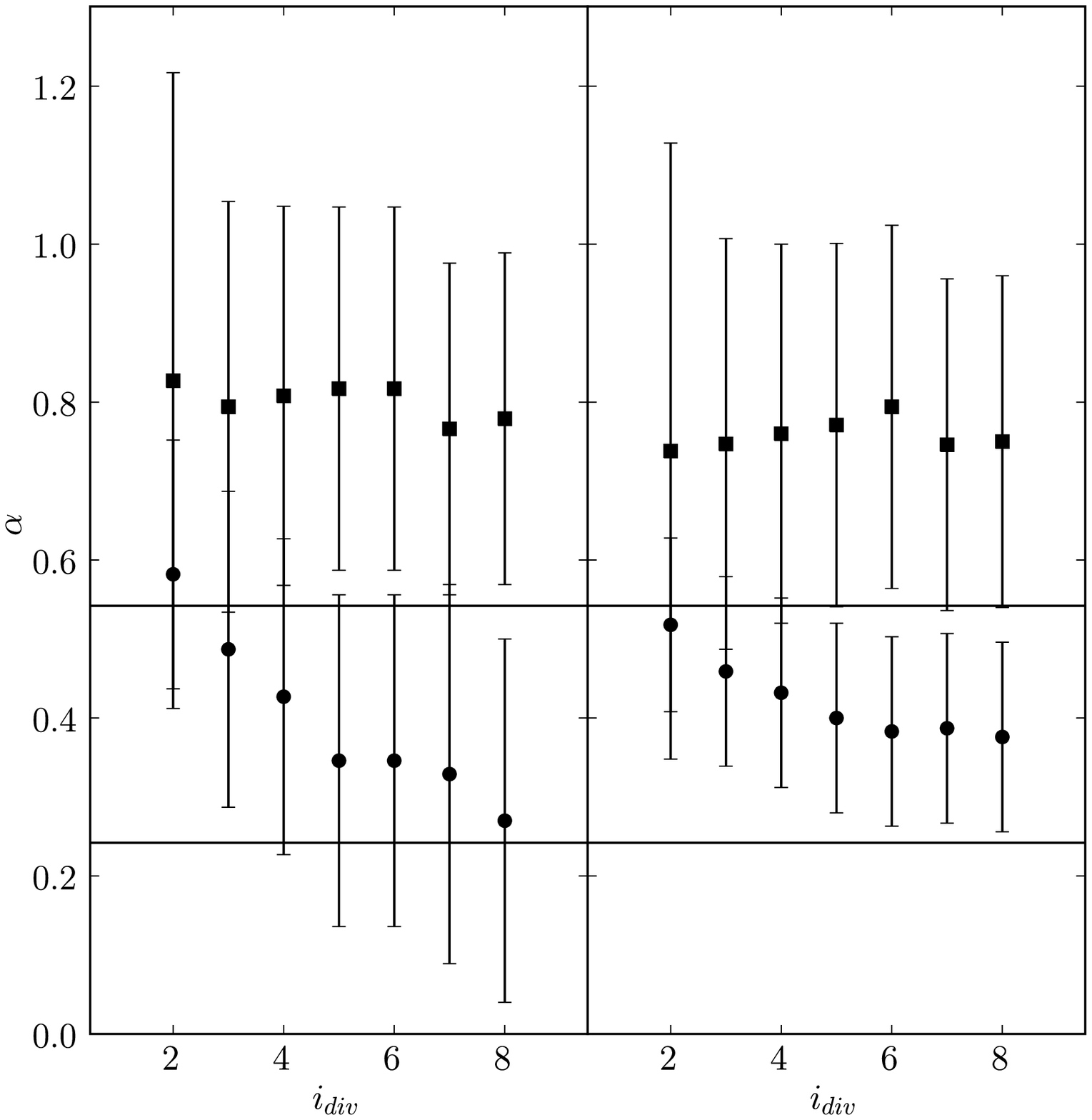}
   \caption{\label{fig:alpha-inc}}
\end{figure}

Fig~\ref{fig:alpha-inc}. Best-fit slopes $\alpha$ of the \Cold (squares) and \Hot (circles) populations versus the inclination division $i_{div}$ separating the two. Left is using only those objects with follow-up. Right is using all observations. The 1-$\sigma$ range in $\alpha_{Close}$ is bracketed by the two horizontal lines.

%%now put the figures

\newpage

\end{document}